\documentclass[mnsc,nonblindrev]{informs3aa} % current default for manuscript submission

\OneAndAHalfSpacedXII % current default line spacing
\usepackage{natbib}
 \bibpunct[, ]{(}{)}{,}{a}{}{,}%
 \def\bibfont{\small}%

\usepackage[algo2e]{algorithm2e}
\SetKwRepeat{Do}{do}{while}%
\SetKwInput{KwInput}{Input}
\SetKwInput{KwOutput}{Output}
\SetKwInput{KwFunction}{Function}
\SetKwInput{KwParameters}{Parameters}

\SetCommentSty{mycommfont}
\usepackage{enumitem}

\usepackage{subcaption}

\allowdisplaybreaks[4]

\newcommand{\ud}{\mathrm d}

\newcommand{\poly}{\mathrm{poly}}

\def\hat{\widehat}

\def\path{\textsf{PATH}}

\TheoremsNumberedThrough     % Preferred (Theorem 1, Lemma 1, Theorem 2)
\EquationsNumberedThrough    % Default: (1), (2), ...
\begin{document}
%%%%%%%%%%%%%%%%

% Outcomment only when entries are known. Otherwise leave as is and 
%   default values will be used.
%\setcounter{page}{1}
%\VOLUME{00}%
%\NO{0}%
%\MONTH{Xxxxx}% (month or a similar seasonal id)
%\YEAR{0000}% e.g., 2005
%\FIRSTPAGE{000}%
%\LASTPAGE{000}%
%\SHORTYEAR{00}% shortened year (two-digit)
%\ISSUE{0000} %
%\LONGFIRSTPAGE{0001} %
%\DOI{10.1287/xxxx.0000.0000}%

% Author's names for the running heads
% Sample depending on the number of authors;
% \RUNAUTHOR{Jones}
% \RUNAUTHOR{Jones and Wilson}
% \RUNAUTHOR{Jones, Miller, and Wilson}
% \RUNAUTHOR{Jones et al.} % for four or more authors
% Enter authors following the given pattern:
%\RUNAUTHOR{}

% Title or shortened title suitable for running heads. Sample:
% \RUNTITLE{Bundling Information Goods of Decreasing Value}
% Enter the (shortened) title:
\RUNTITLE{Near-Linear Time Local Polynomial Nonparametric Estimation}

% Full title. Sample:
\TITLE{Near-Linear Time Local Polynomial Nonparametric Estimation with Box Kernels}
% Enter the full title:
%\TITLE{}

% Block of authors and their affiliations starts here:
% NOTE: Authors with same affiliation, if the order of authors allows, 
%   should be entered in ONE field, separated by a comma. 
%   \EMAIL field can be repeated if more than one author
\ARTICLEAUTHORS{%
\AUTHOR{Yining Wang}
\AFF{Warrington College of Business, University of Florida, Gainesville, FL 32611, USA.}
\AUTHOR{Yi Wu}
\AFF{Institute of Interdisciplinary Information Sciences, Tsinghua University, Beijing, 100084, China.}
\AUTHOR{Simon S. Du}
\AFF{Paul G. Allen School of Computer Science and Engineering, University of Washington, WA 98195, USA.}
% Enter all authors
} % end of the block

\ABSTRACT{%
	Local polynomial regression \citep{fan1996local} is an important class of methods
	for nonparametric density estimation and regression problems.
	However, straightforward implementation of local polynomial regression has quadratic time complexity which hinders its applicability in large-scale data analysis.
	In this paper, we significantly accelerate the computation of local polynomial estimates
	by novel applications of multi-dimensional binary indexed trees \citep{fenwick1994new}.
	Both time and space complexity of our proposed algorithm is nearly linear in the number of input data points.
	Simulation results confirm the efficiency and effectiveness of our proposed approach.
}%

% Sample 
%\KEYWORDS{deterministic inventory theory; infinite linear programming duality; 
%  existence of optimal policies; semi-Markov decision process; cyclic schedule}

% Fill in data. If unknown, outcomment the field
%\HISTORY{}

\maketitle
%%%%%%%%%%%%%%%%%%%%%%%%%%%%%%%%%%%%%%%%%%%%%%%%%%%%%%%%%%%%%%%%%%%%%%
%
%Nonparametric density and function estimation is an important question in both statistics and machine learning research \citep{larry2006all,tsybakov2009introduction,friedman2001elements}.
%Example applications of nonparametric function estimation include smoothing and prediction of econometric trends like loan management, market profit prediction and wheat crop predictions \citep{gyorfi2006distribution}. 
%The nonparametric density estimation problem, on the other hand, is useful for exploratory analysis of unlabeled data,
%and also has applications in other fields such as computer vision \citep{miller2003practical} and computational fluid mechanics \citep{eugeciouglu2000efficient}.
%%Example applications include ...

\section{Introduction}

Big data analytics has become essential for modern operations research and operations management applications \citep{choi2018big,hazen2018back,mivsic2019data}.
Statistics methods, such as nonparametric density and function estimation \citep{larry2006all,tsybakov2009introduction,friedman2001elements},
play important roles in the prediction of econometric trends like loan management and market profit prediction \citep{gyorfi2006distribution},
as well as exploratory data analysis as preliminary steps to operations management problems, such as the understanding and estimation of customers' demand or preferences in
revenue management applications.
Naturally, big data scenarios pose unique challenges from \emph{computing} aspects, as many classical statistical computational routines (especially nonparametric statistical methods)
are computationally heavy and not scalable to large amounts of data. 

The focus of this paper is on nonparametric density estimation and regression problems.
Let $\mathcal X\subseteq\mathbb R^d$ be a compact domain in $\mathbb R^d$,
which is conventionally taken to be the unit cube $[0,1]^d$ for convenience.
In the \emph{nonparametric density estimation} problem, $n$ independent samples $X_1,\cdots,X_n\in\mathcal X$
are obtained as
\begin{equation}
	X_1,\cdots,X_n \;\overset{i.i.d.}{\sim}\; f_0,
\end{equation}
where $f_0$ is the density of an unknown distribution $P_0$ supported on $\mathcal X$.
In the \emph{nonparametric regression} problem, on each of $n$ ``design points'' $X_1,\cdots,X_n\in\mathcal X$
a ``response'' or ``measurement'' $Y_i$ is obtained according to the model
\begin{equation}
	Y_i = m_0(X_i) + \xi_i,\;\;\;\;\;i=1,\cdots n,
\end{equation}
where $m_0:\mathcal X\to\mathbb R$ is an unknown regression model of interest,
and $\{\xi_i\}$ are independent sub-Gaussian noise variables satisfying $\mathbb E[\xi_i|X_i] = 0$.
The objective in both nonparametric estimation problems is to construct estimates $\hat f_n$ or $\hat m_n$ such that 
the \emph{mean-square error (MSE)} 
\begin{equation*}
	\int_{\mathcal X}\big|\hat f_n(x)-f_0(x)\big|^2\ud x \;\;\text{or}\;\; 
	\int_{\mathcal X}\big|\hat m_n(x)-m_0(x)\big|^2\ud x
\end{equation*}
is minimized.

%One important aspect of the nonparametric estimation problems is the \emph{minimal assumptions} imposed on the unknown models of interest $f_0$ or $m_0$.
%More specifically, only \emph{smoothness} type assumptions such as bounded high-order derivatives of $f_0$ or $m_0$ are imposed,
%which contrasts classical \emph{parametric} approaches that formulate $f_0$ or $m_0$ as linear or generalized linear models.
%This allows the unknown function to belong to large and comprehensive function classes, 
%which cover most functions that arise in practical applications.

\begin{figure*}[t!]
	\centering
	\begin{subfigure}[t]{0.45\textwidth}
		\includegraphics[width=\textwidth]{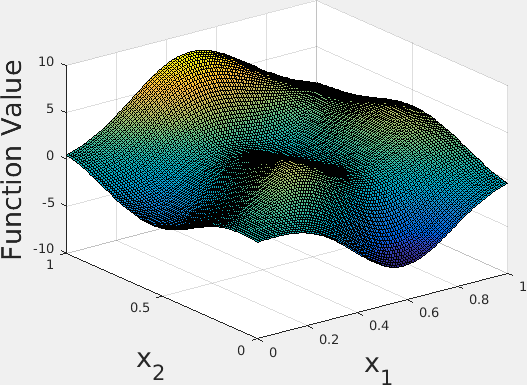}
		\caption{The ground truth function.}
	\end{subfigure}	
	\quad
	\begin{subfigure}[t]{0.45\textwidth}
		\includegraphics[width=\textwidth]{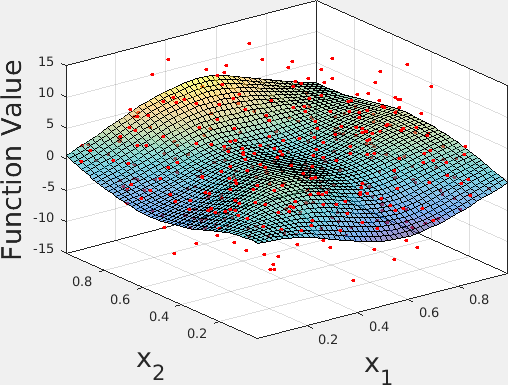}
		\caption{Naive impl., $n=4000$, 75 secs}
	\end{subfigure}
	\quad
	\begin{subfigure}[t]{0.45\textwidth}
		\includegraphics[width=\textwidth]{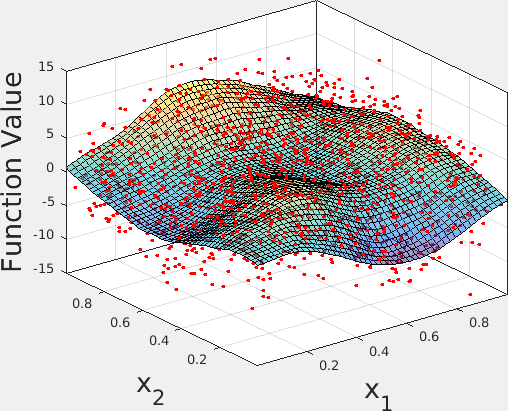}
		\caption{Fast impl., $n=16000$, 10 secs}
	\end{subfigure}
	\caption{Illustration of local polynomial regression with naive and fast implementation for fitting a two dimensional smooth function
		{$f(x,y)=3(1-x)^2e^{-x^2-(y+1)^2} - 10(0.2x - x^3 - y^5)e^{-x^2-y^2} - e^{-(x+1)^2-y^2}/3$.}
		Red points are observations (training points), down sampled $10\times$ for better visualization.
		%For presentation purpose we only show $10\%$ training points.
		%As number of training points increases, the fitted function is closer to the ground truth.
		The number of testing points is large ($\sim 1,000,000$), as reconstruction of the entire function is desired.
		Under such settings, the naive implementation (middle panel) takes $75$ secs to process $n=4000$ observations (training points), % and processing $16000$ training data takes  $~300s$.
		while our fast implementation (Algorithm~\ref{alg:main}, right panel) only takes $10$ secs to process $n=16000$ observations.
		As a result, there is a visible difference between the fitted curves under the naive and the fast implementation,
		because the naive implementation can only process a small number of observations under given time budget and therefore needs to significantly ``over-smooth'' the data,
		leading to inaccurate function reconstruction.
		Indeed, the mean-square errors of the naive and fast implementation are $9\times 10^{-3}$ and $2\times 10^{-3}$ respectively, which is a near $5\times$ gap.
	}
	\label{fig:2d_illustration}
\end{figure*}

The nonparametric density estimation and regression problems have a long history of study, dating back to the 1920s \citep{whittaker1922new}.
A large family of methods have been developed and their properties analyzed,
including kernel smoothing \citep{friedman2001elements,gyorfi2006distribution},
spline smoothing \citep{reinsch1967smoothing,geer2000empirical,green1993nonparametric},
wavelet smoothing \citep{donoho1998minimax,donoho1994ideal,hardle2012wavelets}
and local polynomial regression \citep{fan1992variable,fan1993local,fan1996local}.

In this paper, we concentrate on the local polynomial regression method introduced in \citep{fan1992variable,fan1993local,fan1996local}.
Compared to its competitors, local polynomial regression has the benefits of being adaptive to non-uniform design densities and regression functions of unbounded values \citep{hastie1993local}.
In Sec.~\ref{sec:lpr} we give a succinct description of these methods and also summarize some of their basic properties.

While the statistical properties of local polynomial regression are well understood, computational efficiency has been less studied.
In particular, straightforward implementation of local polynomial regression typically has \emph{quadratic} time complexity $O(sn)$
to calculate $\hat f_n(z_i)$ or $\hat m_n(z_i)$ on  $s$ ``testing points'' $z_1,\cdots,z_s\in\mathcal X$,
which is prohibitively slow for analysis of large data sets.
We give an illustrative example in Figure \ref{fig:2d_illustration}, which shows that the computational bottleneck of the naive implementation of local polynomial regression
significantly restricts the number of observations (training points) it processes under given time budget, leading to inaccurate function estimates.

The apparent difficulties of computationally efficient local polynomial regresstion motivate the following question:
\begin{question}
	Given $n$ training data points $\{(X_i,Y_i)\}_{i=1}^n$ and $s$ testing points $z_1,\cdots,z_s\in\mathcal X$,
	can we compute local polynomial estimates $\hat f_n(z_1),\cdots,\hat f_n(z_s)$ or $\hat m_n(z_1),\cdots,\hat m_n(z_s)$ in 
	$O((n+s)\poly\log n)$ time?
	\label{ques:main}
\end{question}
Essentially, Question \ref{ques:main} concerns estimation algorithms whose time complexity is \emph{nearly linear} in the total number of training and testing points $n+s$,
apart from possible poly-logarithmic factors.
On the other hand, a linear dependency on $n+s$ is clearly necessary as the number of inputs is already on the order of $n+s$.

We give an affirmative answer to Question \ref{ques:main} in this paper by accelerating local polynomial regression
using efficient data structures.
In particular, applying a multi-dimensional extension of binary indexed trees \citep{fenwick1994new}
we are able to reduce the computation time on each testing point from $O(n)$ to $O(\log^d n)$.
Furthermore, to avoid space complexity growing exponentially with dimension $d$,
we consider a \emph{lazy allocation} strategy that allocates memory on the fly with the help of a Hash table,
which achieves near-linear space complexity.
Our computations of all local polynomial estimates are exact, and therefore all statistical properties of local polynomial regression
are automatically preserved.

Finally, we remark that certain existing nonparametric estimation methods can also achieve near-linear time complexity,
such as the B-spline \citep{de1978practical} and/or falling factorial basis \citep{wang2014falling} construction in smoothing splines
and block thresholding of wavelet coefficients \citep{cai1999adaptive}.
However, both smoothing spline and wavelet thresholding (shrinkage) methods become very complicated for multi-dimensional and non-uniformly spaced
data points.
For example, the extension of smoothing splines to higher dimension requires complicated constructions of the ``thin-plate'' splines,
and in the case of wavelet smoothing the basis functions are complicated and difficult to compute when design points are not evenly spaced.
On the other hand, the local polynomial regression method easily handles both cases of multi-dimensionality
and unevenly spaced design points without much additional complexity.
% Appendix here
% Options are (1) APPENDIX (with or without general title) or 
%             (2) APPENDICES (if it has more than one unrelated sections)
% Outcomment the appropriate case if necessary
%
% \begin{APPENDIX}{<Title of the Appendix>}
% \end{APPENDIX}
%
%   or 
%
% \begin{APPENDICES}
% \section{<Title of Section A>}
% \section{<Title of Section B>}
% etc
% \end{APPENDICES}

% References here (outcomment the appropriate case) 

\section{Local polynomial regression}\label{sec:lpr}

In this section we give a succinct description of local polynomial regression \citep{fan1996local} in the context of nonparametric estimation problems,
with high-level summarization of its properties.

Let $k\in\mathbb N$ be the desired polynomial degree used in estimation.
Practical choices of $k$ are small non-negative integers, such as $k=0$ (local mean averaging),
$k=1$ (local linear regression) and $k=2$ (local quadratic regression).
For any $z\in\mathbb R^d$, define the polynomial mapping $\psi_z:\mathbb R^d\to\mathbb R^D$,
$D=1+d+\cdots+d^k$ as
\begin{align*}
	\psi_z(x) &:= [1, x_1-z_1,\cdots,x_d-z_d,(x_1-z_1)^2, (x_1-z_1)(x_2-z_2), \cdots, (x_d-z_d)^2,\\&\cdots, (x_1-z_1)^k, \cdots, (x_d-z_d)^k].
\end{align*}
At a higher level, the polynomial mapping $\psi_x(\cdot)$ is an expansion of basis of degree-$k$ polynomials on $\mathbb R^d$,
appropriately offset so that $\psi_z(z) = [1,0,\cdots,0]$.
As we shall see immediately, the mapping $\psi_z$ plays a fundamental role in local polynomial regression methods.

%We are now ready to give formal descriptions of the local polynomial regression estimate.
We first consider the regression problem $Y_i=m_0(X_i)+\xi_i$.
We start by introducing the concepts of \emph{kernel functions} and \emph{bandwidths}, which are crucial to almost every nonparametric estimation method:
\begin{enumerate}
	\item A \emph{kernel function} $K:\mathbb R\to\mathbb R$ that satisfies constraints $\int_{-\infty}^\infty K(u)\ud u = 1$, $\int_{-\infty}^\infty K(u)u\ud u=0$ and 
	$\int_{-\infty}^\infty K^2(u)\ud u < \infty$ is used to ``average'' neighboring data points in a certain manner.
	Common kernel functions include the box kernel $K(u)=\mathbb I[|u|\leq 1/2]$, the Gaussian kernel $K(u)=\frac{1}{\sqrt{2\pi}}e^{-u^2/2}$ and the 
	Epanechnikov kernel $K(u)=\frac{3}{4\sqrt{5}}\max(0, 1-u^2/5)$.
	
	\item A \emph{bandwidth} parameter $h>0$ is used to control the size of the ``neighborhood'' of a certain test point $z\in\mathcal X$
	in which training points are incorporated and smoothed.
	A small bandwidth $h$ will have fewer effective training points rendering the variance of the resulting estimate large,
	while a large bandwidth $h$ tends to ``over-smooth'' the unknown function over a large domain
	and therefore leading to larger estimation bias.
\end{enumerate}

To estimate the value of the unknown regression function $m_0$ at a specific testing point $z$, first solve
a weighted least-squares problem
\footnote{Although in the literature the $\ell_2$ norm $\|z-X_i\|_2=\sqrt{\sum_{j=1}^d(z_j-X_{ij})^2}$ is more commonly used,
	replacing the $\ell_2$ norm with $\ell_\infty$ does not affect the statistical properties of $\hat\theta_n$,
	because $\ell_2$ and $\ell_\infty$ are equivalent norms when dimension $d$ is fixed.}
\begin{equation}
	\hat\theta_n = \arg\min_{\theta\in\mathbb R^D} \sum_{i=1}^n{(Y_i-\theta^\top\psi_z(X_i))^2 K\left(\frac{\|z-X_i\|_\infty}{h}\right)}
	\label{eq:lpr}
\end{equation}
where $\|z-X_i\|_\infty = \max_{1\leq j\leq d}|z_j-X_{ij}|$.
Afterwards, $\hat m_n(z)$ is taken to be $\hat\theta_n^\top\psi_z(z)$, which corresponds to the first component of $\hat\theta_n$.

\begin{example}
	When $K(\cdot)$ is taken to be the box kernel $K(u)=\mathbb I[|u|\leq 1/2]$, 
	Eq.~(\ref{eq:lpr}) can be re-formulated as
	$$
	\hat g_n = \arg\min_{g\in\mathcal P_k} \sum_{X_i\in B_h(z)} (Y_i-g(X_i))^2
	$$
	where $\mathcal P_k$ is the class of all degree-$k$ polynomials on $\mathbb R^d$ and $B_h(z)=\{u\in\mathcal X: \|u-z\|_\infty\leq h/2\}$
	denotes the neighborhood of $z$ with radius $h$.
	The re-formulation of $\hat g_n$ is a clear interpretation of the polynomial fitting (in $\mathcal P_k$) and the local properties (in $B_h(z)$).
\end{example}

The density estimation problem is slightly more involved than local polynomial regression.
Here we adopt the approach in \citep{cattaneo2017simple} by re-casting the density estimation problem as a regression problem.
For every $X_i\in\mathcal X$ define empirical cumulative density function (CDF)
\begin{equation}
	\hat F(X_i) := \frac{1}{n}\sum_{j=1}^n\mathbb I\left[X_{j1}\leq X_{i1}, \cdots, X_{jd}\leq X_{id}\right]
\end{equation}
that approximates the true CDF $F(X_i) = \int_{\mathcal X}\mathbb I[u_1\leq X_{i1},\cdots,u_d\leq X_{id}]f_0(u)\ud u$.
Afterwards, a local polynomial fit of $\hat F$ is computed as
\begin{equation*}
	\hat\beta_n = \arg\min_{\beta\in\mathbb R^D} \sum_{i=1}^n{(\hat F(X_i)-\theta^\top\psi_z(X_i))^2 K\left(\frac{\|z-X_i\|_\infty}{h}\right)}.
\end{equation*}
Because $f_0 = \partial^d F/\partial x_1\cdots\partial x_d$,
an estimate $\hat f_n(z)$ can be obtained by reading off the component in $\hat\beta_n$ corresponding to $(z_1-x_1)(z_2-x_2)\cdots(z_d-x_d)$
multiplied by $d!$.
Note that $k\geq d$ is required for this density estimation method.

Local polynomial regression enjoys a wide range of desired properties.
When bandwidths $h$ are properly tuned (for example by cross-validation), 
local polynomial regression is minimax optimal when the underlying model $f_0$ or $m_0$
belongs to H\"{o}lder or Sobolev smoothness classes.
In addition, when bandwidths are selected \emph{locally}, the resulting estimation is optimal even for spatially heterogeneous functions \citep{lepski1997optimal}.
The local polynomial regression estimator also adapts to non-uniform and non-smooth design densities and boundaries \citep{fan1992variable,fan1996local,cheng1997automatic,hastie1993local},
properties that do not hold for kernel smoothing estimators.

\section{Our method}

We describe our proposed method for local polynomial regression with the box kernel that achieves
$O((n+s)\log^d n)$ time complexity and $O(n\log^d n)$ space complexity,
which is the main contribution of this paper.

We start by formulating sufficient statistics required to compute local polynomial estimates,
and giving a simple nearly linear-time algorithm for the univariate case $d=1$ as a warm-up exercise.
We then proceed with a formal discretization argument for higher dimensions, and explain how binary indexed trees
can be applied to enable fast calculations.
Finally, a lazy memory allocation scheme is employed to avoid memory explosion in high-dimensional binary indexed trees.

\subsection{Sufficient statistics of local polynomial regression}

When the box kernel $K(u)=\mathbb I[|u|\leq 1/2]$ is used, 
the local polynomial regression formulation in Eq.~(\ref{eq:lpr}) is an ordinary least squares (OLS)
problem on $\mathcal B(z,h) := \{(X_i,Y_i): \|X_i-z\|_\infty\leq h/2\}$, where $z\in\mathcal X$ is the testing point.
The solution $\hat\theta_n$ admits the following closed form:
\begin{equation}
	\hat\theta_n = \left[\sum_{\mathcal B(z,h)} \psi_z(X_i)\psi_z(X_i)^\top\right]^{-1}\left[\sum_{\mathcal B(z,h)}Y_i\psi_z(X_i)\right].
	\label{eq:ols}
\end{equation}

Recall that $\psi_z(X_i)$ is a multivariate polynomial function in $X_i$. 
The sufficient statistics required to compute Eq.~(\ref{eq:ols}) can then be organized as follows:
\begin{enumerate}
	\item $\sum_{(X_i,Y_i)\in \mathcal B(z,h)} X_{i1}^{j_1}X_{i2}^{j_2}\cdots X_{id}^{j_d}$ for $j_1+\cdots+j_d\leq 2k$;
	\item $\sum_{(X_i,Y_i)\in\mathcal B(z,h)} Y_i X_{i1}^{j_1}\cdots X_{id}^{j_d}$ for $j_1+\cdots+j_d\leq k$.
\end{enumerate}
{
Let $M$ be the total number of terms summarized above.
For each $\ell\in\{1,2,\cdots,M\}$, define $T_\ell(X_i,T_i)$ to be either $X_{i1}^{j_1}X_{i2}^{j_2}\cdots X_{id}^{j_d}$
or $Y_i X_{i1}^{j_1}\cdots X_{id}^{j_d}$, where $j_1+\cdots+j_d\leq 2k$ or $j_1+\cdots+j_d\leq k$ are indices corresponding to term $\ell\leq M$.
The terms $T_\ell(X_i,Y_i)$ for $\ell\leq M$ are then \emph{sufficient statistics} required to compute the local polynomial estimate $\hat\theta_n$
in Eq.~(\ref{eq:ols}).
}

Once the sufficient statistics $\mathcal T_\ell(X_i,Y_i)$, $(X_i,Y_i)\in\mathcal B(z,h)$ are accumulated for a specific testing point $z\in\mathcal X$ and bandwidth $h>0$,
the estimate $\hat\theta_n$ and subsequently $\hat m_n(z)$ or $\hat f_n(z)$ can be computed in $O(D^3)$ time and $O(D^2)$ space
via Eq.~(\ref{eq:ols}), both treated as constants under fixed-dimension ($d$) settings.

Thus, the computational question reduces to efficient computation of sufficient statistics for every testing point $z$ and bandwidth $h$.
\footnote{In this paper we consider constant bandwidths $h$ for every testing point. Nevertheless, our algorithm framework can be 
easily modified to handle heterogeneous bandwidths as well.}
To simplify our presentation, we abstract the sufficient statistics as
\begin{equation}
	\mathcal T_\ell^{z,h} = \sum_{\mathcal B(z,h)} T_\ell(X_i,Y_i), \;\;\;\;\;\;\ell=1,2,\cdots,M,
\end{equation}
where $M$ is the total number of sufficient statistics required.
A simple brute-force algorithm loops over all $\{(X_i,Y_i)\}_{i=1}^n$ and takes $O(n)$ time to calculate $\mathcal T_\ell^{z,h}$ for each testing point $z$,
which results in an overall $O(sn)$ quadratic time complexity.
Further optimization of computational procedures of $\mathcal T_\ell^{z,h}$ is the focus of the rest of this section.

\subsection{The univariate case: a warm-up exercise}

Before giving the full description of our algorithm, we first consider the basic univariate case ($d=1$) and show
a simple algorithm that computes $\mathcal T_\ell^{z,h}$ for each $z\in\mathcal X$ and $h>0$ in $O(\log n)$ time.
Though simple, the univariate algorithm serves as a conceptual starting point for our follow-up development of general high-dimensional algorithms.

Let $\{(X_i,Y_i)\}_{i=1}^n$ be the training data points where $X_i,Y_i\in\mathbb R$.
As pre-processing steps, $(X_i,Y_i)$ are sorted in ascending order with respect to $X_i$, such that $X_1\leq X_2\leq\cdots\leq X_n$,
and for each $\ell=1,\cdots,M$ the cumulative statistics
\begin{equation}
	\mathcal A_\ell(i) = \sum_{j\leq i}T_\ell(X_j,Y_j) \;\;\;\;\;\;i=1,2,\cdots,n
\end{equation}
are computed by a sweeping algorithm from $i=1$ to $i=n$.
It is clear that the pre-processing can be done in $O(n\log n+Mn)$ time with $O(Mn)$ space.

Afterwards, for each testing point $z\in\mathbb R$ and $h>0$, the left and right ``boundaries''
$L = \arg\max\{i: X_i<z-h/2\}$ and $U=\arg\max\{i: X_i\leq z+h/2\}$ are found by binary searches.
The sufficient statistic $\mathcal T_\ell^{z,h}=\sum_{z-h/2\leq X_i\leq z+h/2}T_\ell(X_i,Y_i)$ can then be computed as
\begin{equation}
	\mathcal T_\ell^{z,h} = \mathcal A_\ell(U) - \mathcal A_\ell(L).
	\label{eq:partial-sum}
\end{equation}
The time required to compute $\hat m_n(z)$ can then be upper bounded by $O(\log n + M)$.
The overall time complexity on $s$ testing points is then $O((n+s)(\log n+M))$,
which is simply $O((n+s)\log n)$ because $M$ only depends on $d$ and $k$, both treated as constants in this paper.

\subsection{Discretization in multiple dimensions}

\begin{figure*}[t]
	\centering
	\includegraphics[width=0.32\textwidth]{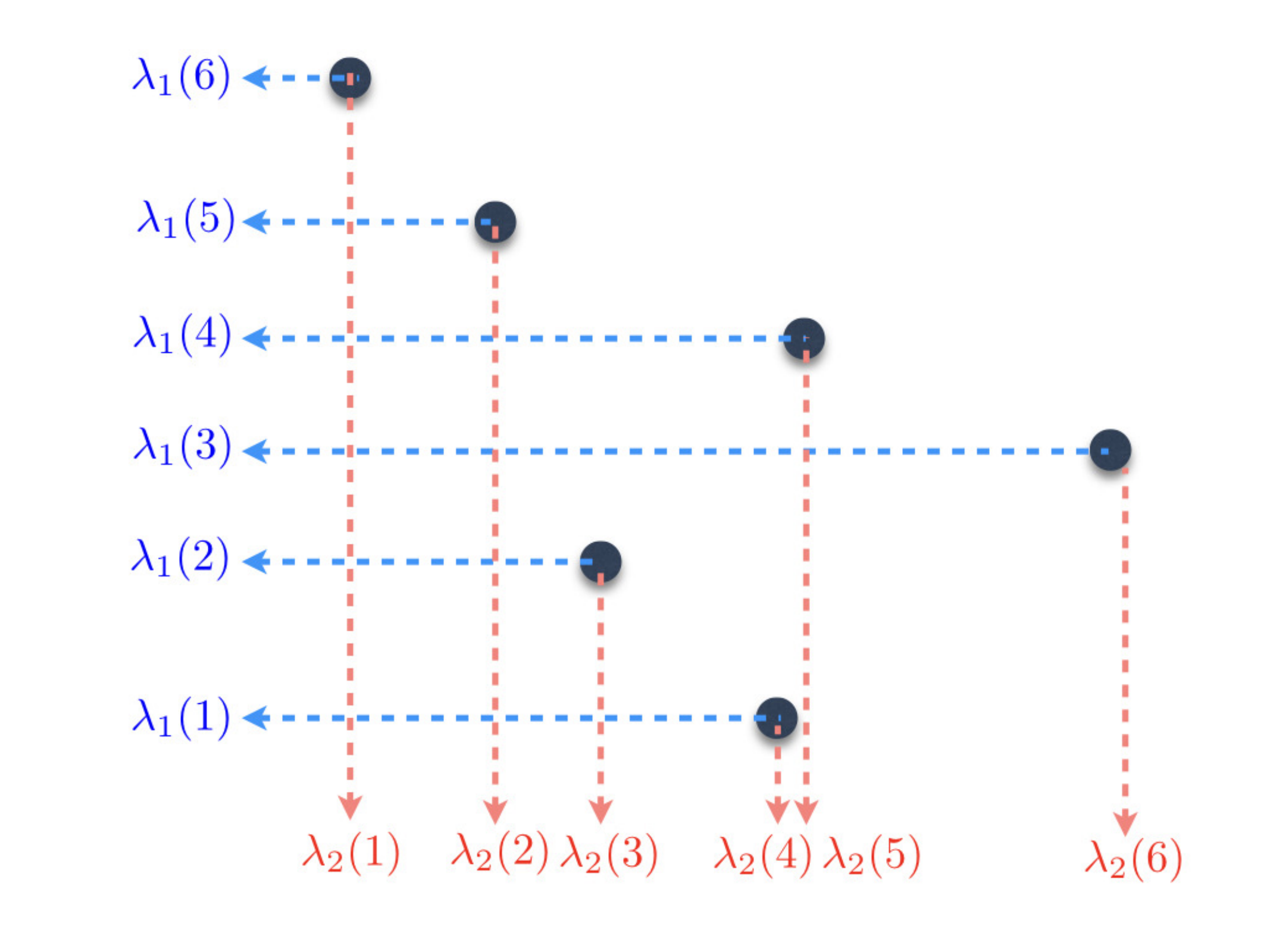}
	\includegraphics[width=0.32\textwidth]{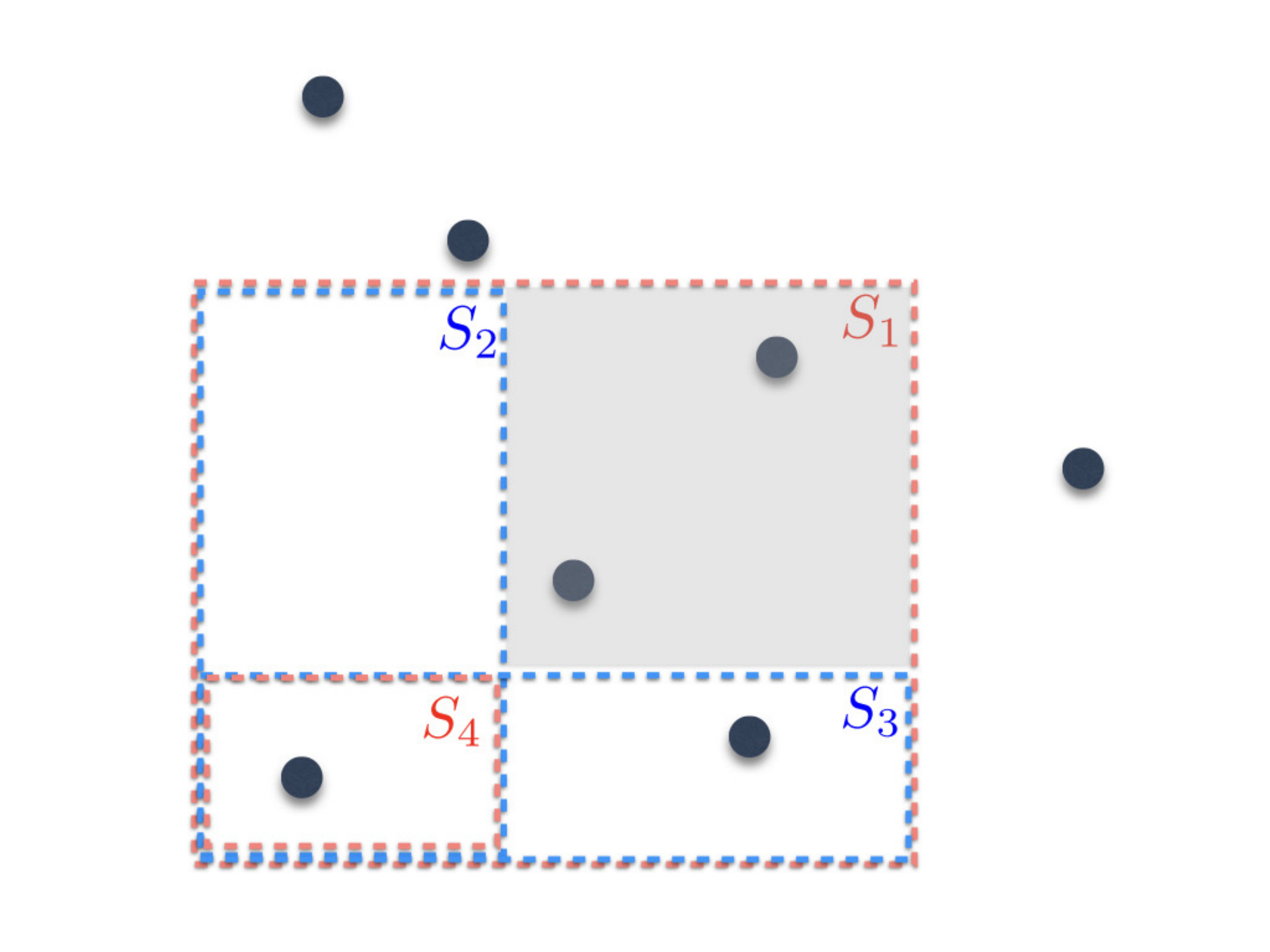}
	\includegraphics[width=0.32\textwidth]{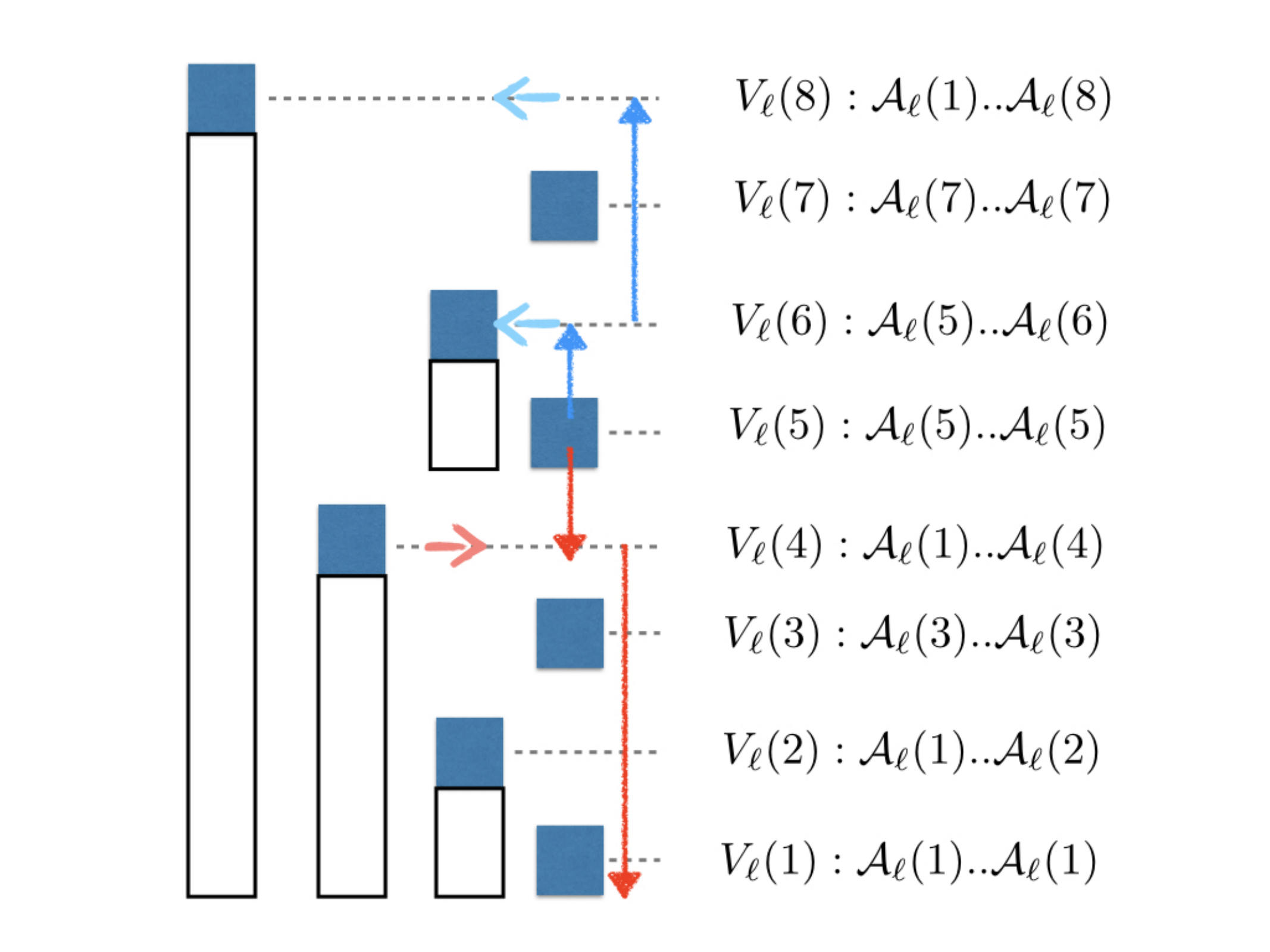}
	\caption{\textbf{The left figure} demonstrates how discretization is carried out when $d=2$, where each of the two dimensions is treated separately
		and $\lambda_j(1)$ through $\lambda_j(6)$ record the values on a particular dimension $j\in\{1,2\}$.
		\textbf{The middle figure} illustrates the inclusion-exclusion principle used to calculate cumulative statistics in the shaded region of interest,
		by considering $S_1-S_2-S_3+S_4$.
		\textbf{The right figure} depicts the one-dimensional binary indexed tree structure for a toy example of $n=8$ data points,
		with each $V_\ell(\cdot)$ indicating over what region is the cumulative statistics supposed to be calculated.
		The red arrows indicate the \emph{interrogation path} when $\mathcal A_\ell(5)$ is evaluated ($5\to 4\to 0$),
		and the blue arrows indicate the \emph{update path} when $T_\ell(5)$ is added ($5\to 6\to 8$).}
	\label{fig:main}
\end{figure*}

The sorting and partial sum idea in Eq.~(\ref{eq:partial-sum}) can be formally generalized to multiple dimensions $d>1$
via the argument of \emph{discretization}. (See the leftmost plot of Figure \ref{fig:main} for a graphical illustration.)
Specifically, for each dimension $j\in\{1,\cdots,d\}$, 
the values $X_{1j}, X_{2j},\cdots,X_{nj}\in\mathbb R$ are sorted in ascending order.
We also denote $\lambda_j(i)$ as the $i$th smallest value in $\{X_{1j},\cdots, X_{nj}\}$ (repetitions counted as multiple values).
By performing such ``discretization'' on all dimensions, 
every training point $X_i\in\mathcal X$ can be mapped to a unique $k$-tuple $\chi_i = (\chi_{i1},\cdots,\chi_{id}) \in [n]^d$,
where $[n]=\{1,2,\cdots,n\}$.

Similar to the definition of $\mathcal A_\ell(i)$ in the $d=1$ setting, for the general multivariate case define
$$
\mathcal A_\ell(i_1,\cdots,i_d) = \sum_{t=1}^n\mathbb I[\chi_{t1}\leq i_1,\cdots,\chi_{td}\leq i_d] T_\ell(X_t,Y_t)
$$
for $i_1,\cdots,i_d\in[n]$.
Suppose for now that $\mathcal A_\ell(i_1,\cdots,i_d)$ are available for all $\ell$ and $i_1,\cdots,i_d$ by some pre-processing procedure.
To compute the estimate $\hat m_n(z)$, first find $L_1,\cdots,L_d$ and $U_1,\cdots,U_d$ by binary searches, where (for $j=1,\cdots,d$)
\begin{align}
	L_j &= \arg\max\left\{i: \lambda_j(i) < z_j-h/2\right\};\nonumber\\
	U_j &= \arg\max\left\{i:\lambda_j(i)\leq z_j+h/2\right\}.\label{eq:luj}
\end{align}
The sufficient statistics $\mathcal T_\ell^{z,h}$ can then be computed as 
$$
\mathcal T_\ell^{z,h} = \sum_{\nu_1,\cdots,\nu_d\in\{0,1\}^d} (-1)^{\nu_1+\cdots+\nu_d}\mathcal A_{\ell}(i_1^{(\nu_1)},\cdots,i_d^{(\nu_d)}),
$$
where $i_j^{(0)}=U_j$ and $i_j^{(1)}=L_j$.
The validity of the computational formula for $\mathcal T_\ell^{z,h}$ is implied by the inclusion-exclusion principle,
and involves $2^d$ evaluations of $\mathcal A_\ell(i_1,\cdots,i_d)$.
An illustrative example of the $d=2$ case is given in the middle plot of Figure \ref{fig:main}.

The question remains to design fast algorithms that compute $\mathcal A_{\ell}(i_1,\cdots,i_d)$
for arbitrary input tuple $(i_1,\cdots,i_d)\in [n]^d$.
In the basic univariate case of $d=1$, this is accomplished by a sweeping pre-processing that records $\mathcal A_\ell(i)$ for all $i\in[n]$.
Such an approach, however, no longer works for $d>1$ because a sweeping algorithm that records all $\mathcal A_\ell(i_1,\cdots,i_d)$
takes $O(n^d)$ space and time complexity and is clearly impractical.
To overcome this difficulty, we consider a data structure named \emph{binary indexed trees} and show how it can 
compute every $\mathcal A_\ell(i_1,\cdots,i_d)$ evaluation in $O(\log^d n)$ time.

\begin{figure}[p]
\scalebox{0.9}{
\begin{algorithm2e}[H]
	\DontPrintSemicolon
	\KwInput{training set $\{(X_i,Y_i)\}_{i=1}^n$, testing set $\{z_i\}_{i=1}^s$, bandwidth $h>0$, dimension $d$, polynomial degree $k$}
	\KwOutput{local polynomial estimates $\{\hat m_n(z_i)\}_{i=1}^s$ or $\{\hat f_n(z_i)\}_{i=1}^s$}
	\BlankLine
	Initialize: hash functions $\hbar_1,\cdots,\hbar_d$ and hash tables $\{\mathcal H_\ell\}_{\ell=1}^M$;\;
	\BlankLine
	$\triangleright$ \emph{pre-processing steps}\;
	Discretization: for every $j\in[d]$ sort $\{X_{ij}\}_{i=1}^n$ in ascending order and label them $\lambda_j(1),\cdots,\lambda_j(d)$;\;
	\For{each $\ell=1,\cdots,M$ and $t=1,\cdots,n$}{
		Compute sufficient statistics $T_\ell(X_t,Y_t$) and find $i_1,\cdots,i_d\in[n]$ such that $\lambda_1(i_1)=X_{t1},\cdots,\lambda_d(i_d)=X_{td}$;\;
		Compute update paths $\path^U(i_1), \cdots, \path^U(i_d)$ using Eq.~(\ref{eq:update});\;
		\For{$i_1'\in\path^U(i_1), i_2'\in\path^U(i_2),\cdots,i_d'\in\path^U(i_d)$}{
			Update: $\mathcal H_\ell(H(i_1',\cdots,i_d')) \gets \mathcal H_\ell(H(i_1',\cdots,i_d')) + T_\ell(X_t,Y_t)$;
		}
	}
	\BlankLine
	$\triangleright$ \emph{compute estimates}\;
	\For{each $t=1,\cdots,s$}{
		Find $L_1,\cdots,L_d$ and $U_1,\cdots,U_d$ defined in Eq.~(\ref{eq:luj}) for $z_t$ and $h$ using binary searches;\;
		\For{each $\ell=1,\cdots, M$}{
			\For{$\nu_1,\nu_2,\cdots,\nu_d\in\{0,1\}$}{
				Initialize $\mathcal A_\ell(i_1^{(\nu_1)},\cdots,i_d^{(\nu_d)})=0$, where $i_j^{(0)}=U_j$ and $i_j^{(1)}=L_j$ for $j\in[d]$;\;
				Compute interrogation paths $\path^I(i_1^{(\nu_1)}), \cdots, \path^I(i_d^{(\nu_d)})$ using Eq.~(\ref{eq:interrogation});\;
				\For{$i_1'\in\path^I(i_1^{(\nu_1)}),\cdots,i_d'\in\path^I(i_d^{\nu_d})$}{
					Accumulate: $\mathcal A_\ell(i_1^{(\nu_1)},\cdots,i_d^{(\nu_d)})\gets \mathcal A_\ell(i_1^{(\nu_1)},\cdots,i_d^{(\nu_d)}) + \mathcal H_\ell(H(i_1',\cdots,i_d'))$;
				}
			}
			Compute $\mathcal T_\ell^{z_t,h} = \sum_{\nu_1,\cdots,\nu_d}(-1)^{\nu_1+\cdots+\nu_d}\mathcal A_\ell(i_1^{(\nu_1)},\cdots,i_d^{(\nu_d)})$, using the inclusion-exclusion principle;\;
		}
		Obtain estimate $\hat m_n(z_t)$ or $\hat f_n(z_t)$ using Eq.~(\ref{eq:ols}) and sufficient statistics $\{\mathcal T_\ell^{z_t,h}\}_{\ell=1}^L$;\;
		}
	\vskip 0.1in
	\caption{The main accelerated local polynomial regression algorithm.}
	\label{alg:main}
\end{algorithm2e}
}
\end{figure}

\subsection{Binary indexed trees}

The \emph{binary indexed tree} or \emph{Fenwick's tree}, invented and first documented by \cite{fenwick1994new},
is a simple yet powerful data structure that supports fast queries and updates of \emph{partial sums},
corresponding to $\mathcal A_\ell(i_1,\cdots,i_d)$ defined in this paper.

In the rightmost panel of Figure \ref{fig:main} we give a toy example with $n=8$ data points and $d=1$.
The $V_\ell(i)$ entries for $i=1,\cdots,n$ are the main data structure kept by the binary indexed tree,
each corresponding to the cumulative statistics of data points in a specific interval.
To query or interrogate a particular partial sum $\mathcal A_\ell(i)$, one starts with $i$ and repeatedly rips off the least significant bit (LSB) in the binary expansion of $i$
and accumulates the entries of $V_\ell$ on the path.
For example, for $i=5$ the query/interrogation path would be $5\to 4\to 0$.
Formally, the following update rule is applied when interrogating partial sum $\mathcal A_\ell(i)$:
\begin{equation}
	\text{\bf Interrogation}: \;\; i \gets i - \textsc{LSB}(i).
	\label{eq:interrogation}
\end{equation}
Here $\textsc{LSB}(i)$ denotes the least significant bit of $i$ in its binary expansion.

To add (update) a data point 
\footnote{Here $ T_\ell(i)$ denotes $T_\ell(X,Y)$ for the data tuple $(X,Y)$ that corresponds to the $i$th position in the sorted data locations.
	$T_\ell(i_1,\cdots,i_d)$ has a similar meaning in higher dimensions.}
$ T_\ell(i)$ to the binary indexed tree,
one has to undo the interrogation path in Eq.~(\ref{eq:interrogation}).
It is easy to verify that the following update procedure is an exact mirroring of the interrogation rule in Eq.~(\ref{eq:interrogation}):
\begin{equation}
	\text{\bf Update}: \;\; i \gets i + \textsc{LSB}(i+1).
	\label{eq:update}
\end{equation}
For example, starting with $i=5$ the updating procedure would take us to $5\to 6\to 8$,
meaning that $V_\ell(5)$, $V_\ell(6)$ and $V_\ell(8)$ are updated with the data entry $T_\ell(5)$ added.

To simplify notations, we write $\path^I(i)$ and $\path^U(i)$ as the interrogation and update paths starting from $i$, respectively.
For example, $\path^I(5) = \{5,4\}$ and $\path^U(5) = \{5,6,8\}$ if $n=8$.
It is a simple observation that for any $i\in[n]$, $|\path^I(i)|, |\path^U(i)| = O(\log n)$.
Thus, both interrogation and updating operations (for $d=1$) can be completed in $O(\log n)$ time.

To extend the univariate binary indexed tree to higher dimensions, consider tuple $i_1,\cdots,i_d\in[n]^d$
at which interrogation or update is needed.
For evaluation of $\mathcal A_\ell(i_1,\cdots,i_d)$, 
accumulate all $V_\ell(i_1',\cdots,i_d')$ such that $i_1'\in\path^I(i_1), i_2'\in\path^I(i_2),\cdots, i_d'\in\path^I(i_d)$
and add them together.
Similarly, when adding a data point $T_\ell(i_1,\cdots,i_d)$, consider all $i_1'\in\path^U(i_1),i_2'\in\path^U(i_2),\cdots,i_d'\in\path^U(i_d)$
and update each $V_\ell(i_1',\cdots,i_d')$ with $T_\ell(i_1,\cdots,i_d)$.
The time complexity for both interrogation and updating is $O(\log^d n)$.

\subsection{Lazy memory allocation via Hashing}

Although the time complexity of $d$-dimensional binary indexed trees is $O(\log^d n)$ and is small,
the space complexity is another story.
A naive implementation of the multi-dimensional binary indexed tree would require $O(n^d)$ memory to 
store all $V_\ell(i_1,\cdots,i_d)$ values.
It is imperative to design more intelligent space allocation algorithms that significantly reduce the $O(n^d)$ space complexity
for practical usages.

An important observation of the multi-dimensional binary indexed tree is that the $V_\ell$ values are extremely \emph{sparse},
with most of the entries equal to zero.
More specifically, with $n$ points in the training data set, the number of non-zero entries of $V_\ell$ is at most $O(n\log^d n)$
because each data point is associated with a update path of length at most $O(\log^d n)$,
which is significantly smaller than $O(n^d)$.
This motivates us to consider a \emph{lazy memory allocation} scheme based on Hash tables,
which would subsequently reduce the space complexity from $O(n^d)$ to nearly $O(n\log^d n)$.

In particular, we construct $M$ hash tables $\{\mathcal H_\ell\}_{\ell=1}^M$ each corresponding to a sufficient statistics $\mathcal T_\ell$ to be maintained.
When adding a new data point $T_\ell(i_1,\cdots,i_d)$, we insert all entries $i_1'\in\path^U(i_1),\cdots,i_d'\in\path^U(i_d)$ into the corresponding Hash table $\mathcal H_\ell$;
similarly when interrogating a point $(i_1,\cdots,i_d)$ we read values off all relevant entries $i_1'\in\path^I(i_1),\cdots,i_d'\in\path^I(i_d)$ from the Hash table $\mathcal H_\ell$ and add them together.

Let $b\in\mathbb N$ be a pre-determined capacity parameter of hash tables $\{\mathcal H_\ell\}_{\ell=1}^M$ to be constructed, which satisfies $b=\Omega(n\log^d n)$. 
Let $\tau\in\mathbb N$ be another pre-defined integer.
Suppose $\hbar_1,\cdots,\hbar_d:[n]\to[b]$ are $\tau d$-wise independent hash functions, meaning that for all $j=1,\cdots,d$ and any $i_1,i_2,\cdots,i_{\tau d}\in[n]$, $a_1,\cdots,a_{\tau d}\in[b]$, 
$$
\Pr\left[\hbar_j(i_1)=a_1\wedge\cdots\wedge \hbar_j(i_{\tau d})=a_{\tau d}\right] = b^{-\tau d}.
$$
A ``composite'' hash function $H:[n]^d\to[b]$ can then be defined as
\begin{equation*}
	H(i_1,\cdots,i_d) := \left(\hbar_1(i_1) + \cdots + \hbar_d(i_d)\right) \mod b.
\end{equation*}
It is a well-known result that $H$ is a $\tau$-wise independent hash function on $[n]^d$ (see for example \citep{pham2013fast,wang2015fast}).
The corresponding entry of $(i_1,\cdots,i_d)\in[n]^d$ in the Hash table $\mathcal H_\ell$ can then be located at
$$
\mathcal H_\ell(H(i_1,\cdots,i_d)).
$$
When collision occurs, standard methods such as \emph{hash chains} or \emph{linear probing} can be employed to resolve collision.
In particular, when $\tau=5$ for linear probing, the expected number of collision would be at a constant level \citep{pagh2009linear,thorup2012tabulation}.

{
\subsection{Summary of time and space complexity}

The capacity of the Hash table constructed in the previous section, $b$, can be set as $b\asymp n\log^d n$,
leading to the expected number of collision to be upper bounded at a constant level.
Hence, the space complexity of the proposed algorithm is $O(n\log^d n)$.

For the time complexity, note that each insertion of a data point $x_i,i\in[n]$ requires $M$ update steps along the update paths in Eq.~(\ref{eq:update}),
and each estimate requires $2^d M$ interrogation steps along the interrogation paths in Eq.~(\ref{eq:interrogation}).
Because $M\leq (2k)^d+k^d$ and the length of each update/interrogation path is upper bounded by $O(\log^d n)$,
the total time complexity of Algorithm \ref{alg:main} is upper bounded by $O((2k)^d (n+s)\log^d n)=O((n+s)\log^d n)$, where both $k$ and $d$
are treated as constants.

}

% and therefore
%the overall probing time is restricted to $O(n\log^d n)$, with a space complexity $b=\Theta(n\log^d n)$.

\section{Numerical results}

\begin{figure*}[t]
	\centering
	\includegraphics[width=0.45\textwidth]{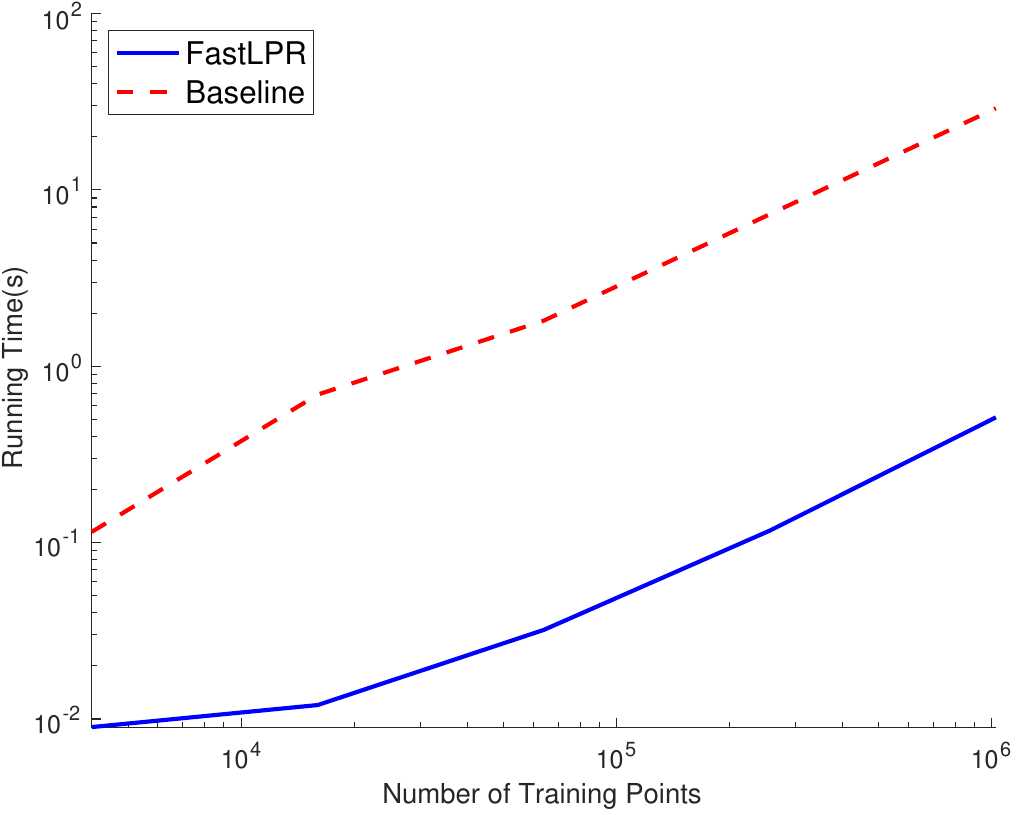}
	\includegraphics[width=0.45\textwidth]{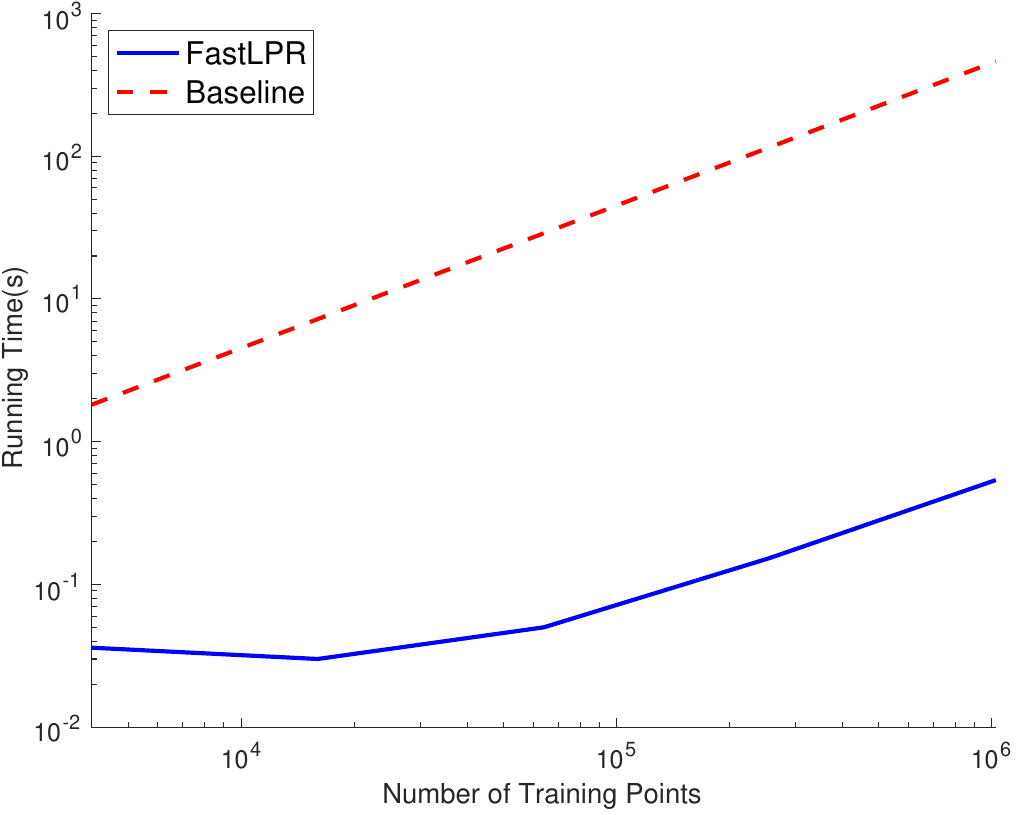}
	\includegraphics[width=0.54\textwidth]{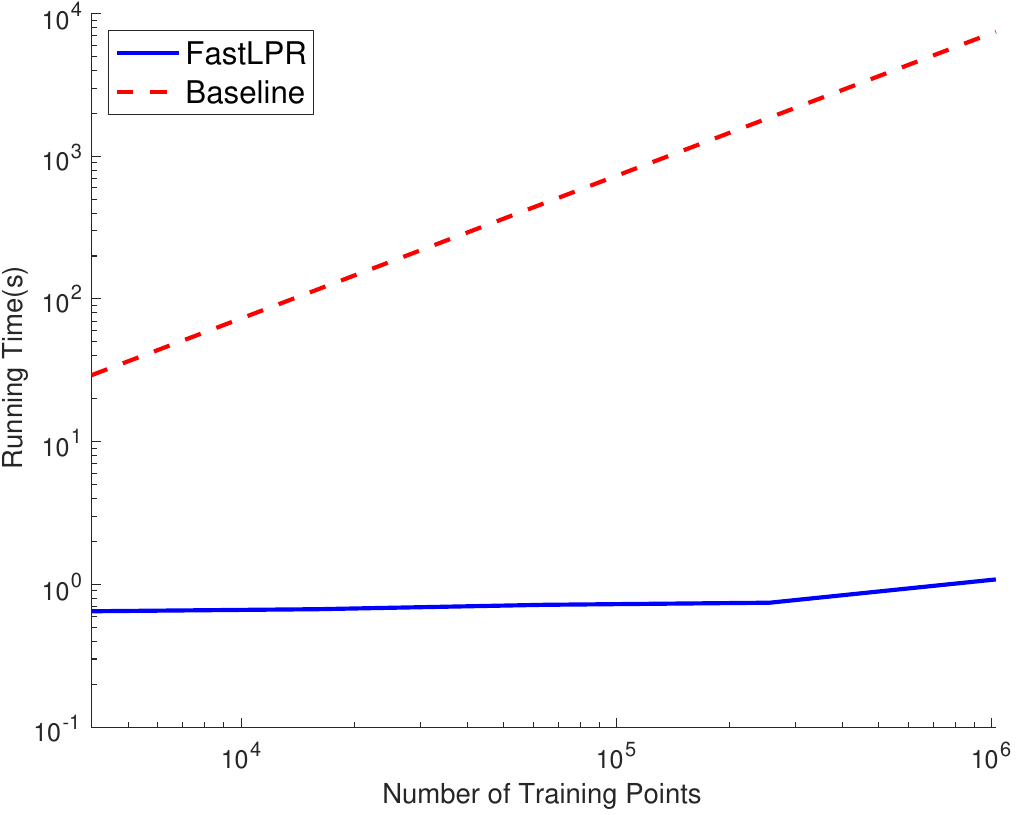}
	\caption{Running times (secs) of fast and baseline implementation of local mean averaging ($k=0$) for univariate data ($d=1$), with varying number of testing points $s=4000$ (left), $s=64000$ (middle) and $s=1024000$ (right).}
	\label{fig:d1k0}
\end{figure*}

\begin{figure*}[t]
	\centering
	\includegraphics[width=0.45\textwidth]{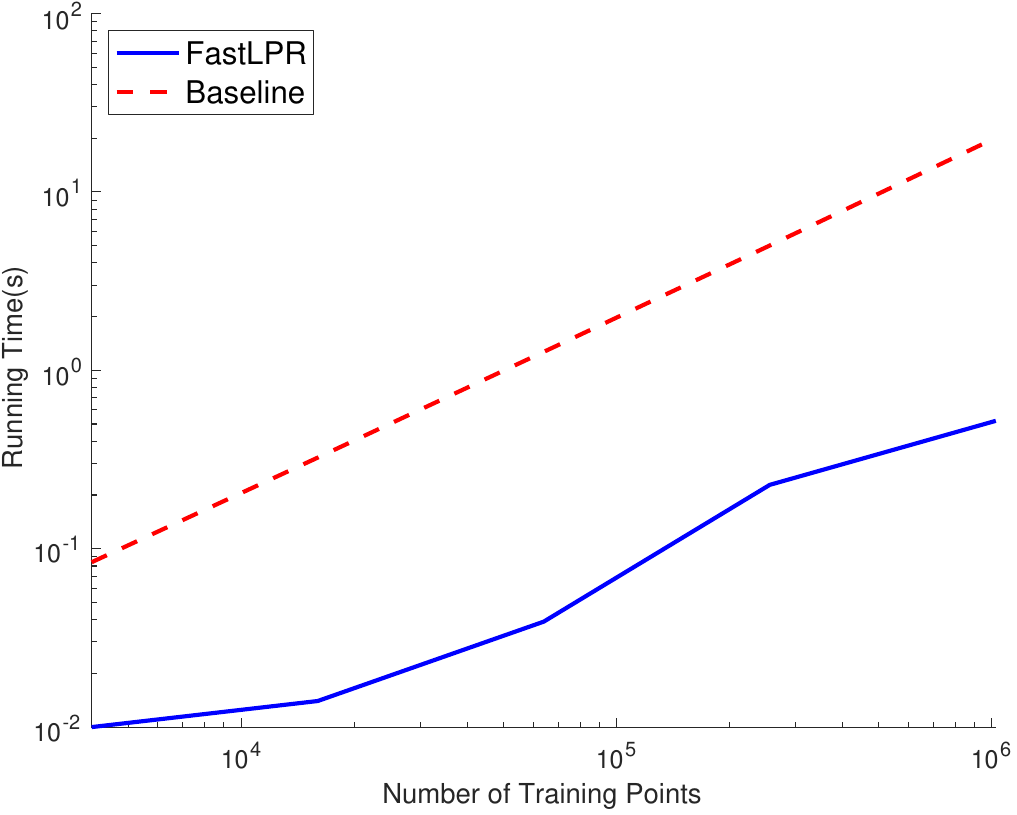}
	\includegraphics[width=0.45\textwidth]{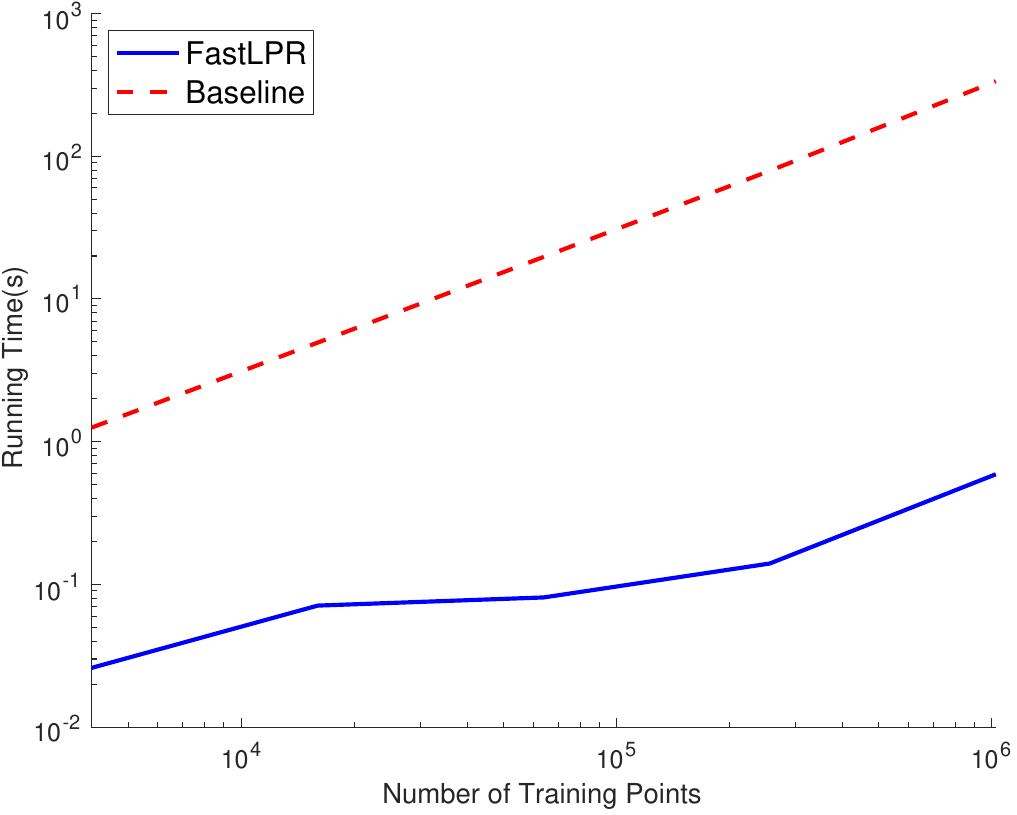}
	\includegraphics[width=0.45\textwidth]{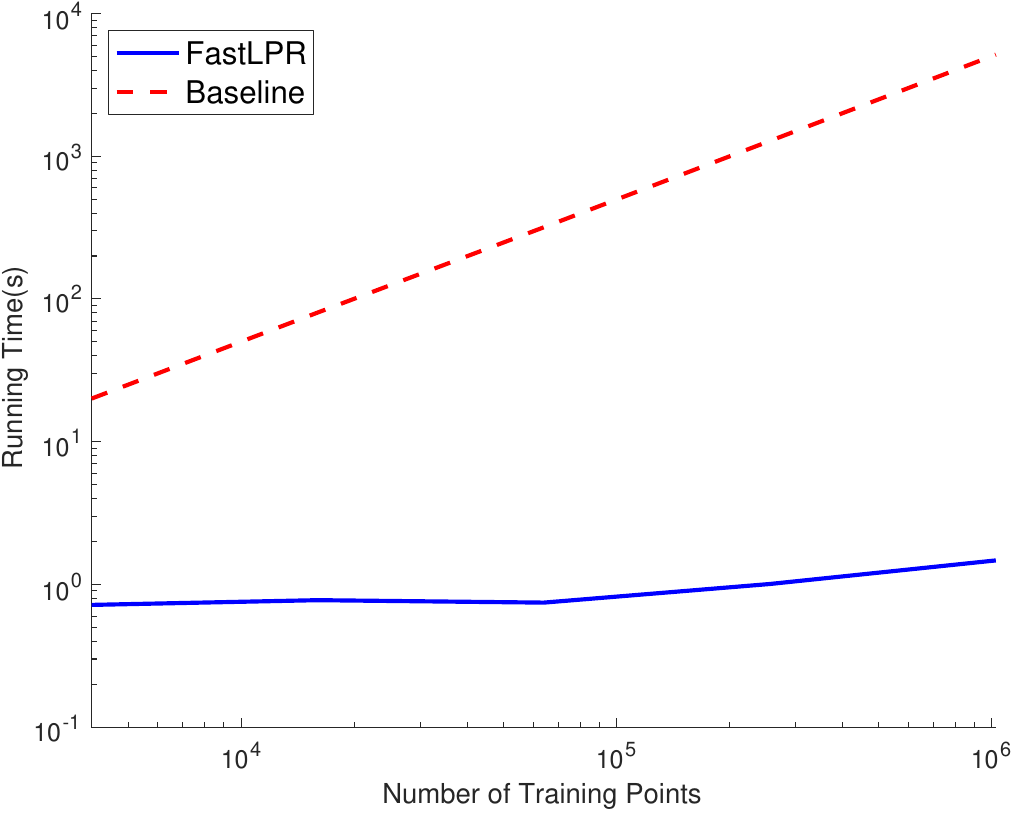}
	\caption{Running times (secs) of fast and baseline implementation of local linear regression ($k=1$) for univariate data ($d=1$), with varying number of testing points $s=4000$ (left), $s=64000$ (middle) and $s=1024000$ (right).}
	\label{fig:d1k1}
\end{figure*}

\begin{figure*}[t!]
	\centering
	\includegraphics[width=0.45\textwidth]{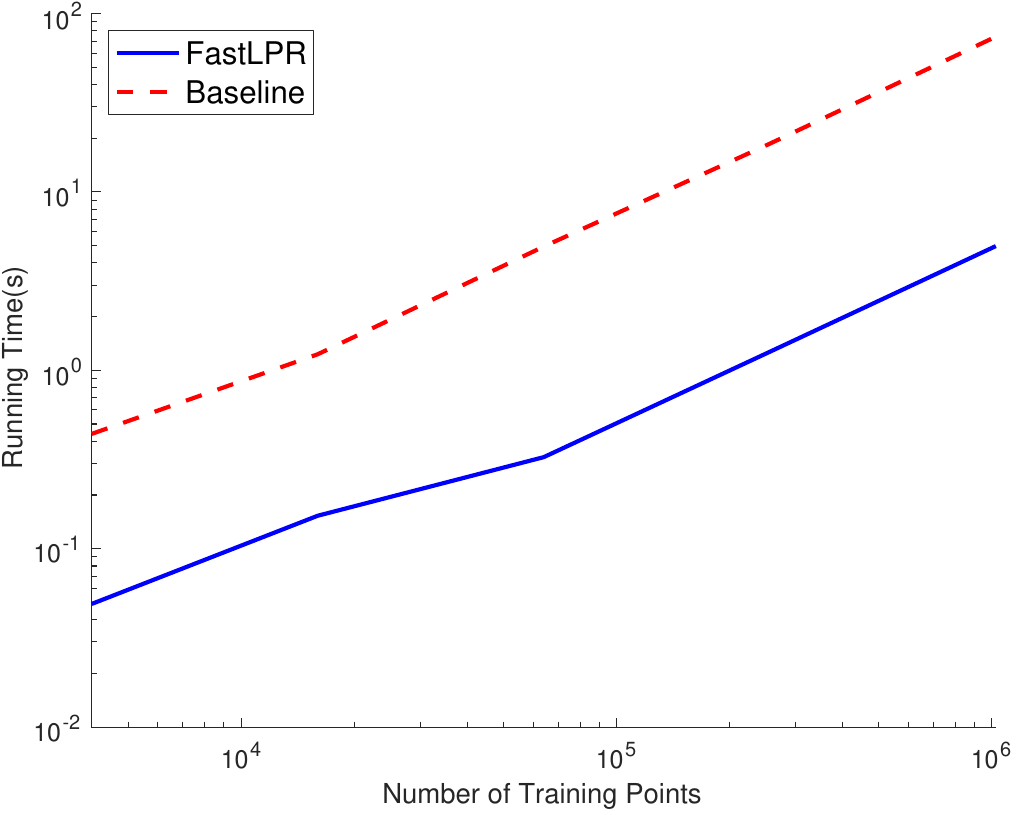}
	\includegraphics[width=0.45\textwidth]{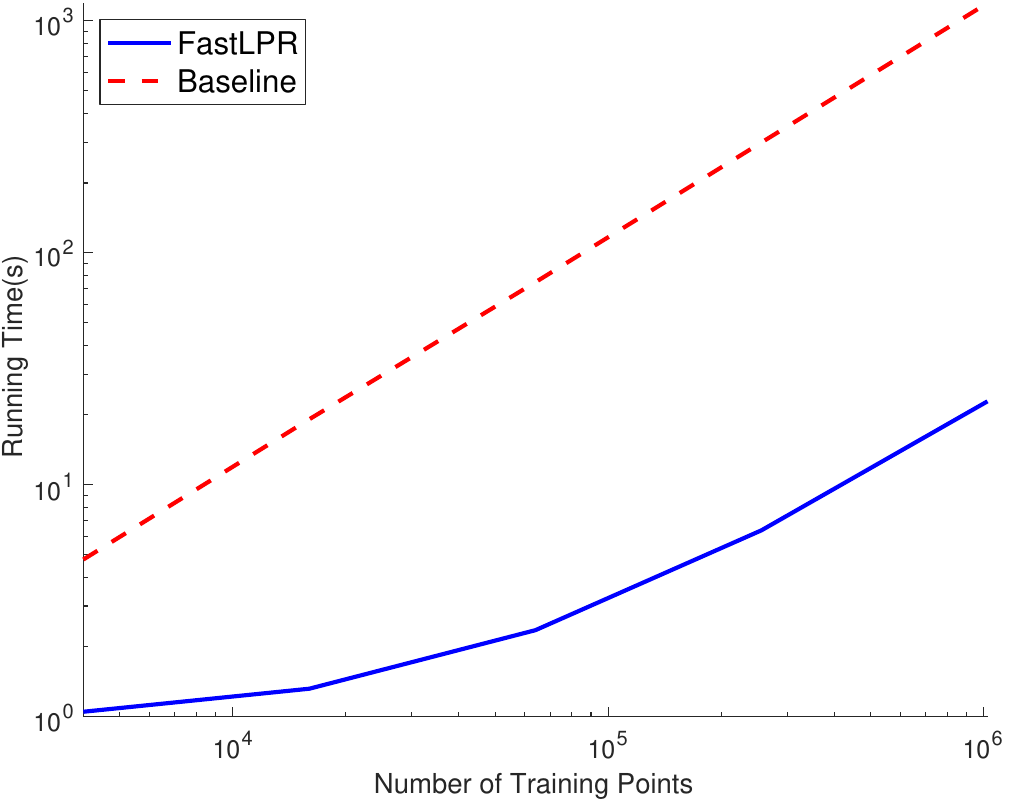}
	\includegraphics[width=0.45\textwidth]{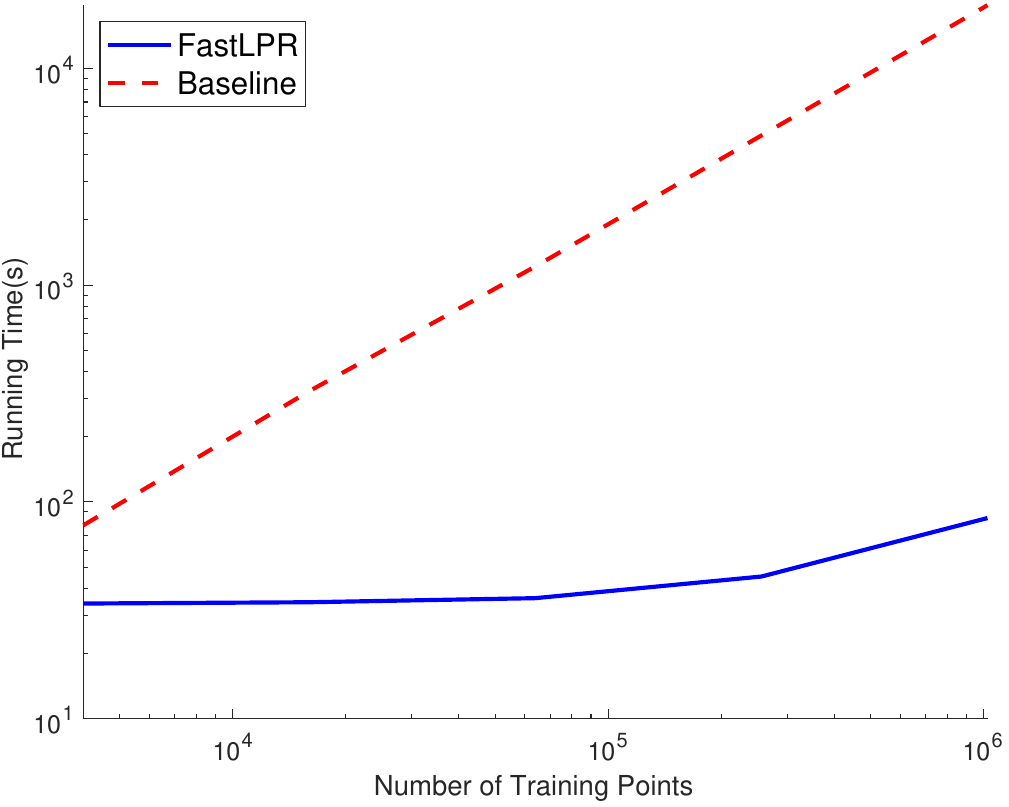}
	\caption{Running times (secs) of fast and baseline implementation of local mean averaging ($k=0$) for bivariate data ($d=2$), with varying number of testing points $s=4000$ (left), $s=64000$ (middle) and $s=1024000$ (right).}
	\label{fig:d2}
\end{figure*}

\begin{figure*}[t!]
	\centering
	\includegraphics[width=0.45\textwidth]{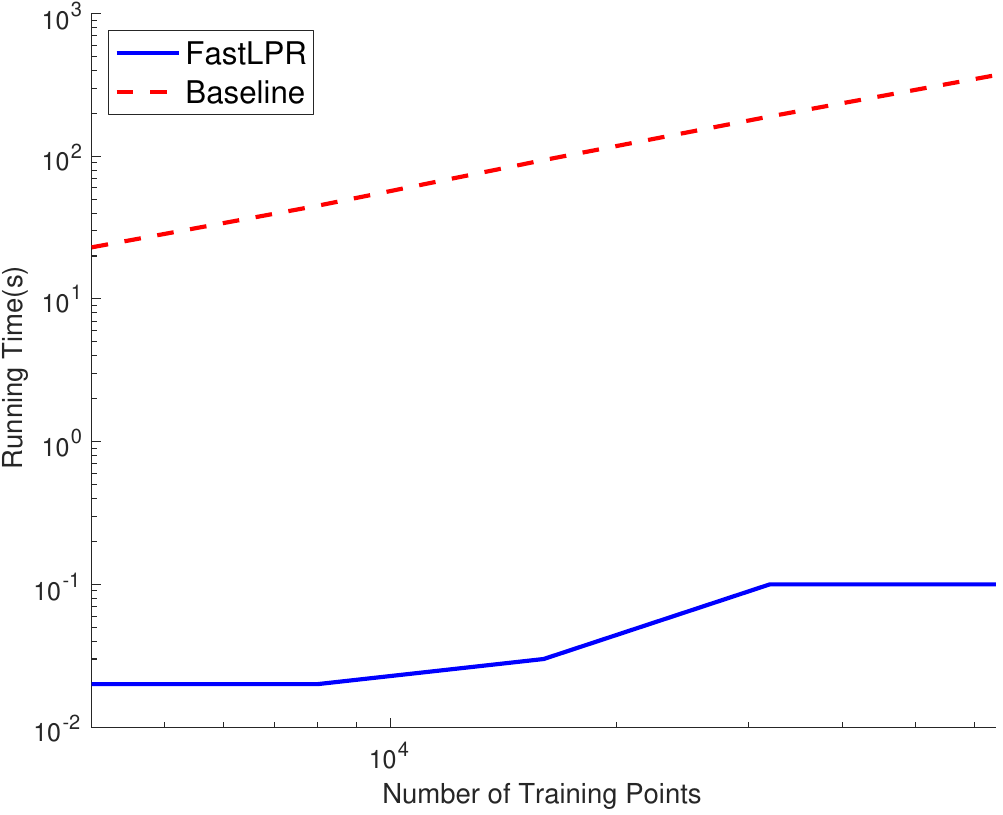}
	\includegraphics[width=0.45\textwidth]{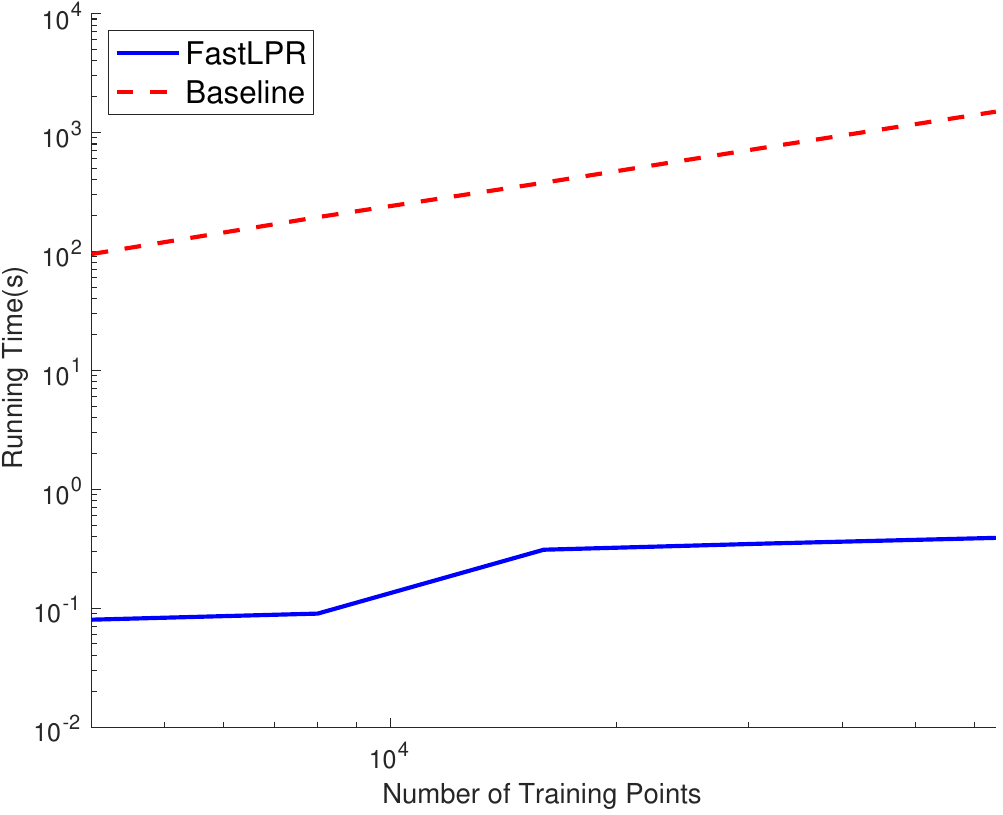}
	\includegraphics[width=0.45\textwidth]{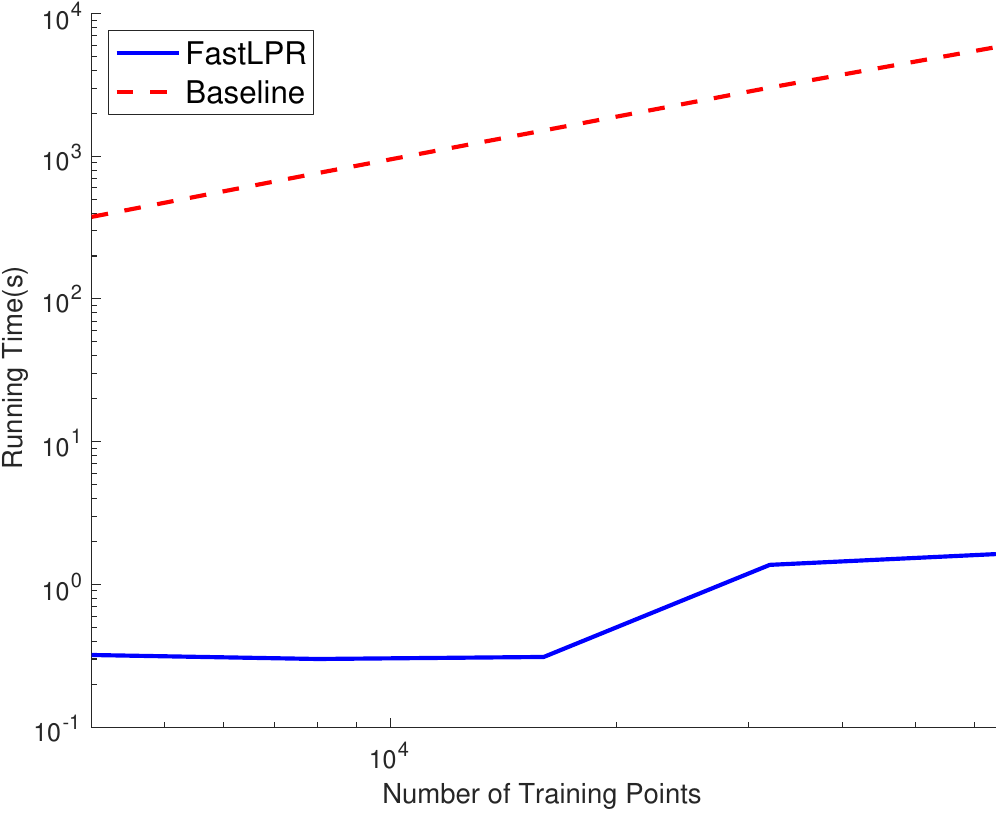}
	\caption{Running times (secs) of fast and baseline implementation of local mean averaging ($k=0$) for bivariate data ($d=3$), with varying number of testing points $s=4000$ (left), $s=16000$ (middle) and $s=64000$ (right).}
	\label{fig:d3k0}
\end{figure*}

\begin{figure*}[t!]
	\centering
	\includegraphics[width=0.45\textwidth]{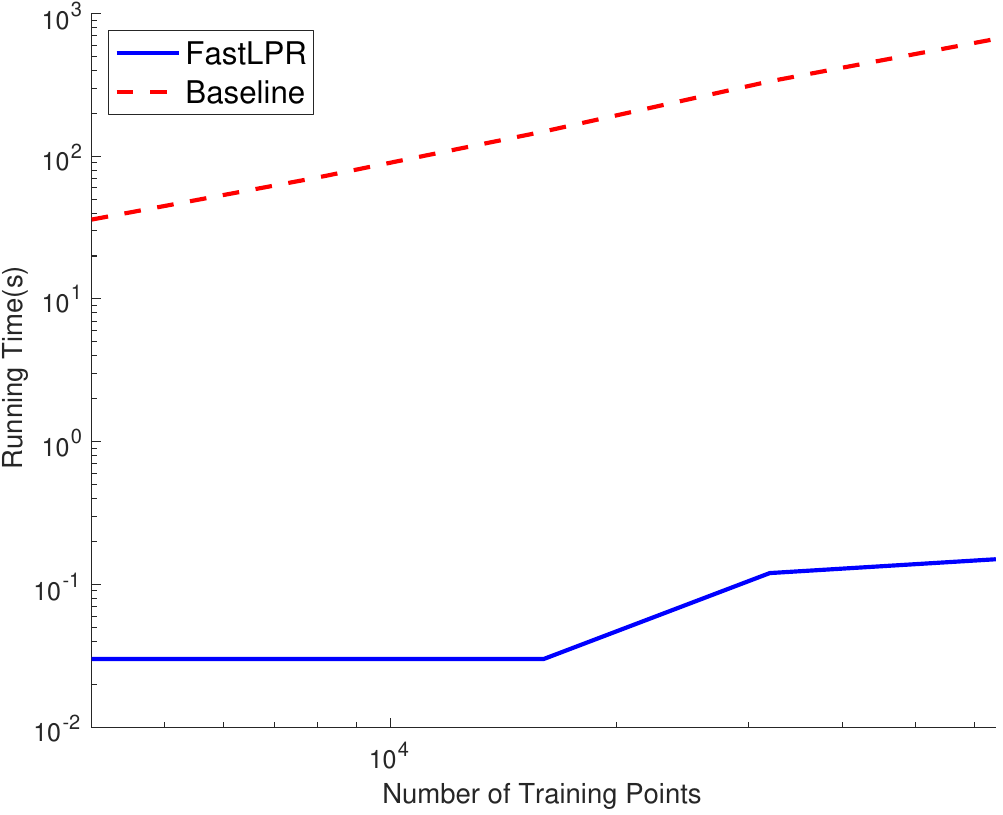}
	\includegraphics[width=0.45\textwidth]{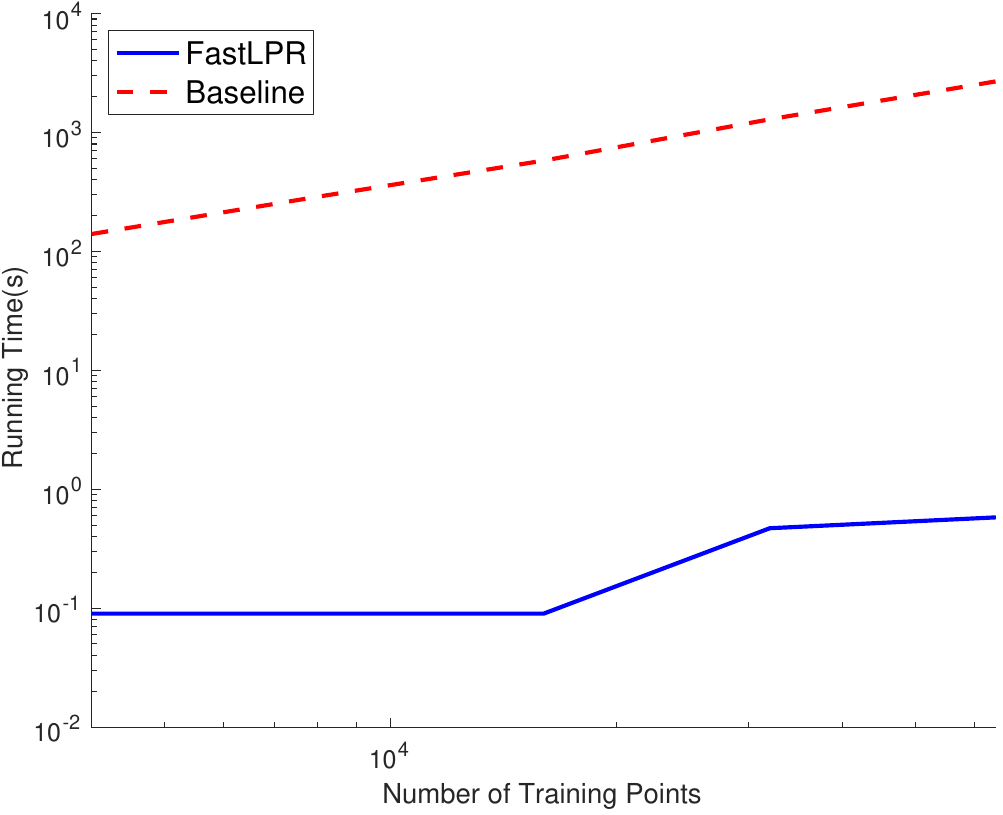}
	\includegraphics[width=0.45\textwidth]{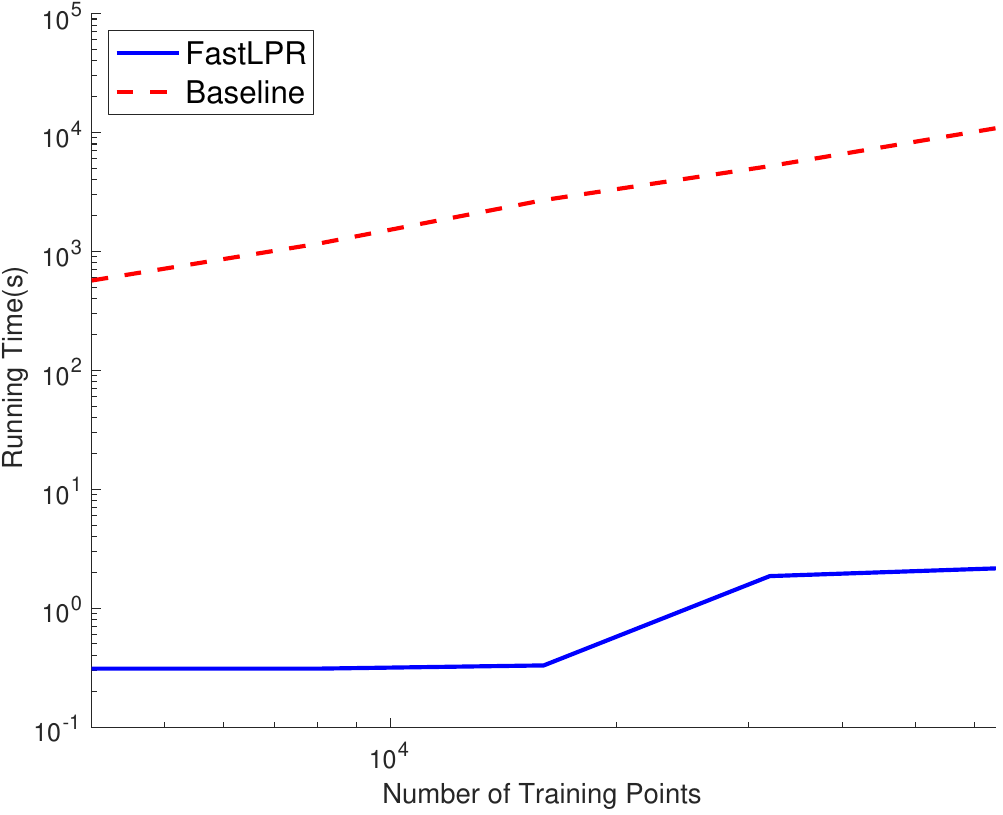}
	\caption{Running times (secs) of fast and baseline implementation of local mean averaging ($k=1$) for bivariate data ($d=3$), with varying number of testing points $s=4000$ (left), $s=16000$ (middle) and $s=64000$ (right).}
	\label{fig:d3k1}
\end{figure*}

\begin{figure*}[t!]
	\centering
	\includegraphics[width=0.45\textwidth]{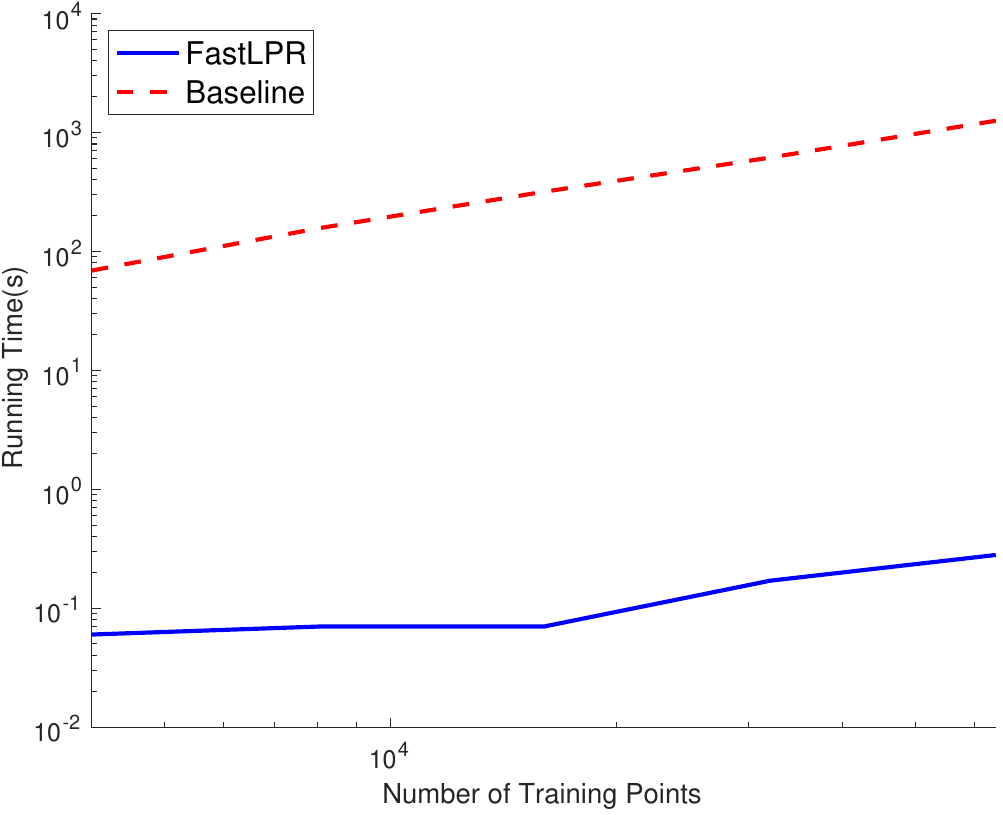}
	\includegraphics[width=0.45\textwidth]{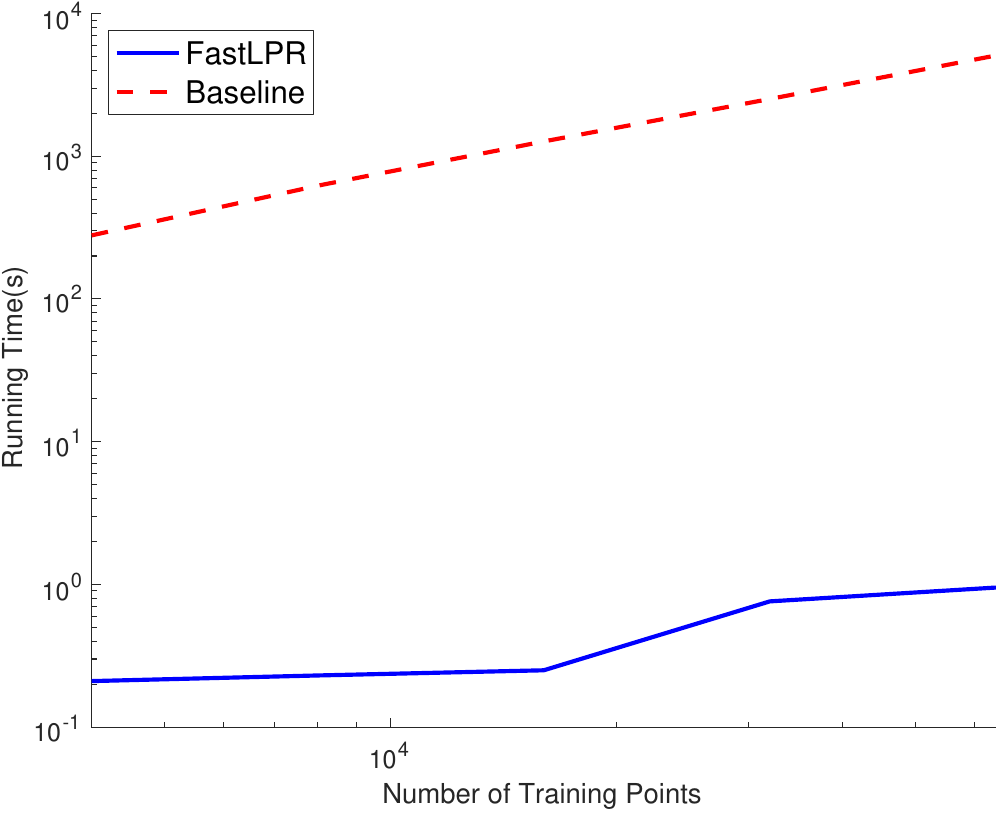}
	\includegraphics[width=0.45\textwidth]{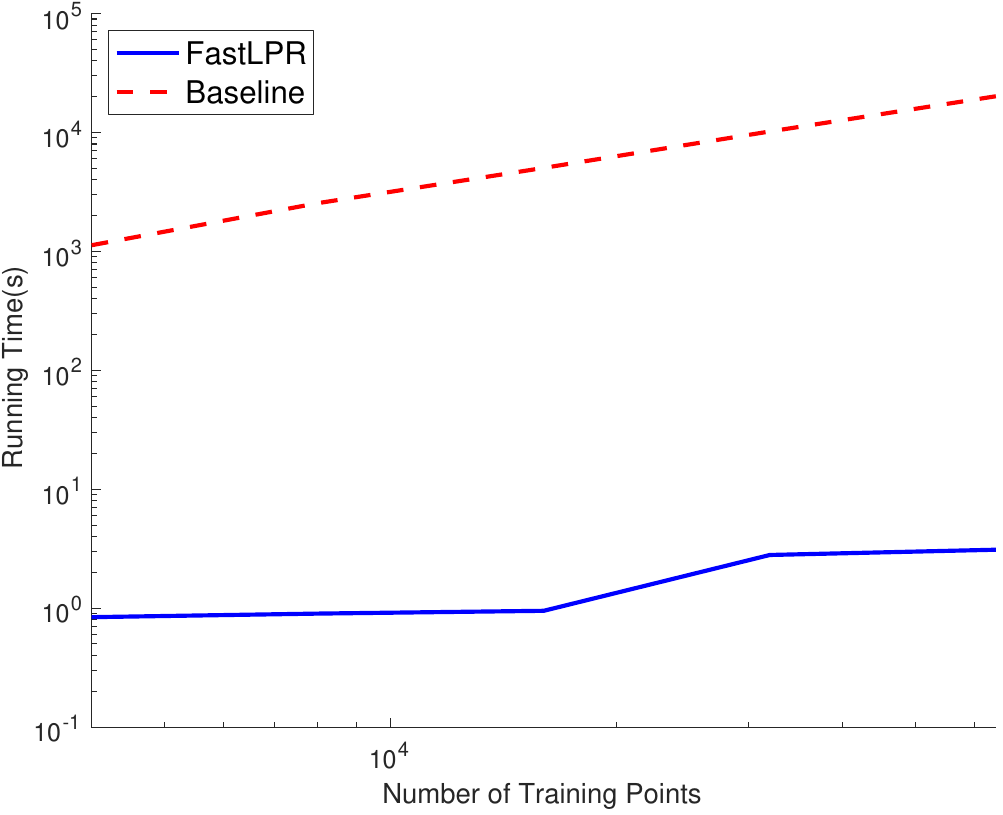}
	\caption{Running times (secs) of fast and baseline implementation of local mean averaging ($k=2$) for bivariate data ($d=3$), with varying number of testing points $s=4000$ (left), $s=16000$ (middle) and $s=64000$ (right).}
	\label{fig:d3k2}
\end{figure*}

In this section we use simulations to verify the effectiveness of our proposed algorithm.
All source codes of our implementation are provided in the supplementary materials accompanying this paper.
The raw data and results in the plotted figures and tables are also provided in a separate online supplement.

%We compare Algorithm~\ref{alg:main} with the standard implementation of local polynomial regression.
%We use the widely used R library \texttt{locpol}\footnote{https://cran.r-project.org/web/packages/locpol} as the baseline, which is implemented in C.
We implement algorithm~\ref{alg:main} using C++.
%The source is code is included in the appendix.
All simulations are done on a computer installed with uBuntu 14.04 LTS 64bit OS, 31.4 Gib Memory, Intel Core i7-4970 CPU @3.60GHz x 8,
except for the running time reported for our own implementation of the brute-force algorithm
which is experimented on a MacBook Pro laptop with macOS Mojave 10.14.6, 16GB memory, 2.3 GHz Intel Core i9
because the former server has been de-serviced at the time this part of the experiment is carried out.

{
Two baselines are considered and compared against our proposed method.
The first baseline is a widely used R library \texttt{locpol}\footnote{https://cran.r-project.org/web/packages/locpol}, which is implemented in C.
A detailed description of the \texttt{locpol} package can be found in \citep{cabrera2012locpol}.
The relevant routines in \texttt{locpol} include \textsf{locCteSmootherC}, \textsf{locLinSmootherC}, \textsf{locCuadSmootherC} and \textsf{locPolSmootherC},
corresponding to local polynomial estimation of $k=0$, $k=1$, $k=2$ and $k>2$, respectively.
Note that for $d\geq 2$ only local constant regression (i.e., $k=0$) is supported in \texttt{locpol}.
While the detailed implementation strategy of \texttt{locpol} is not available, judging from the high computational costs in numerical experiments
we conjecture that the naive brute-force implementation is used which computes $\hat\theta_n$ separately for each data point.

The second baseline is our own implementation of the brute-force computation of local polynomial estimates,
which serves as a second baseline for comparison.
We implement the brute-force approach using C++, which supports general data dimension $d$ and polynomial degree $k$.

}

\begin{table*}
	\centering
	\resizebox{0.7\columnwidth}{!}{%
		\begin{tabular}{ |c|c|c|}
			\hline
			 & Our implementation (MB) & Baseline (MB) \\
			 \hline
			      $k=0,n=4000$ & 72 &  1.3        \\
			      \hline
			      $k=0, n=8000$ & 170 & 1.4\\
			\hline
			      $k=0, n=16000$ & 432& 2.3\\
			      \hline
$k=0, n=32000$ & 1044 & 3.7\\
\hline
$k=0, n=64000$ &2485 & 6.3\\
\hline
$k=1, n=4000$ & 73 & 1.3\\
\hline
$k=1, n=8000$ & 181 & 1.6\\
\hline
$k=1, n=16000$ & 443 & 2.5\\
\hline
$k=1, n=32000$ & 2221 & 3.9\\
\hline
$k=1, n=64000$ & 4945 & 6.6\\
\hline
$k=2, n=4000$ & 354  & 4.2 \\
\hline
$k=2, n=8000$ & 897 & 8.5 \\
\hline
$k=2, n=16000$ &2217  & 16.5\\
\hline
$k=2, n=32000$ & 5280  & 28.4 \\
\hline
$k=2, n=64000$ & 7642& 52.8\\
\hline
		\end{tabular}
	}
	\caption{Comparison of memory consumption between our method and the naive method.
		For all experiments we fix $d=3$ and $s = 4000$.
		\label{tab:memory}
	}
\end{table*}

%Numerical results to be placed here. Explain implementation (language), comparative baselines and experimental settings.
%\textcolor{red}{Do we need to add more implementation details?}

%For each plot, we fix number testing points, polynomial degree and dimension and vary the number of training points.
%In Figure~\ref{fig:s4000}-Figure~\ref{fig:s1024000} we compare Algorithm~\ref{alg:main} (FastLPR) with the naive implementation of local polynomial regression (Baseline).
%In are settings, our algorithm is faster then the baseline.
We first compare the running times of our fast local polynomial regression implementation (Algorithm \ref{alg:main}, denoted as \texttt{FastLPR})
with the baseline R package \texttt{locpol} under various simulation settings in Figure \ref{fig:d1k0} ($d=1,k=0$),
Figure \ref{fig:d1k1} ($d=1,k=1$) and Figure \ref{fig:d2} ($d=2,k=0$).
{
The ground-truth functions $f:\mathbb R^d\to\mathbb R$ are defined as $f(x_1,\cdots,x_d) = \sum_{i=1}^d \sin(x_i)$.}
Note that for $d=2$ we only experiment with local mean averaging ($k=0$) because the \texttt{locpol} package only supports $k=0$ for $d>1$.
For all experimental setups, the running times are reported under varying number of training and testing points,
all uniformly sampled from the unit cube $[0,1]^d$.

\begin{figure*}[t!]
	\centering
	\begin{subfigure}[t]{0.45\textwidth}
		\includegraphics[width=\textwidth]{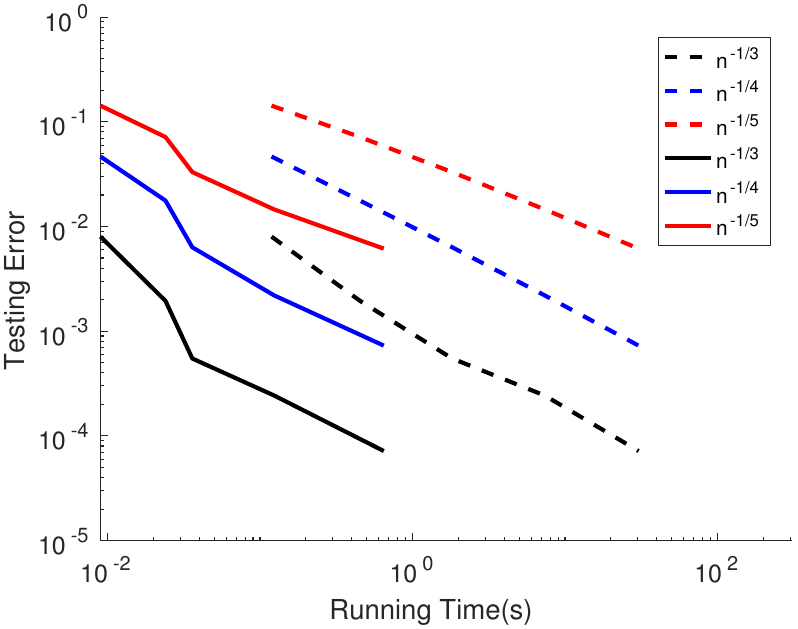}
		\caption{local constant regression ($k=0$)}
	\end{subfigure}	
	\quad
	\begin{subfigure}[t]{0.45\textwidth}
		\includegraphics[width=\textwidth]{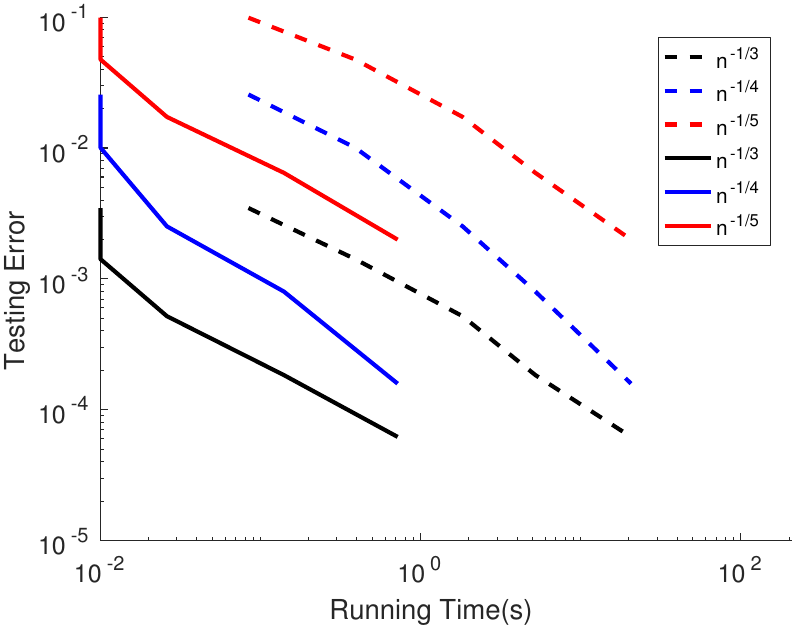}
		\caption{local linear regression ($k=1$)}
	\end{subfigure}
	\quad
	\begin{subfigure}[t]{0.45\textwidth}
		\includegraphics[width=\textwidth]{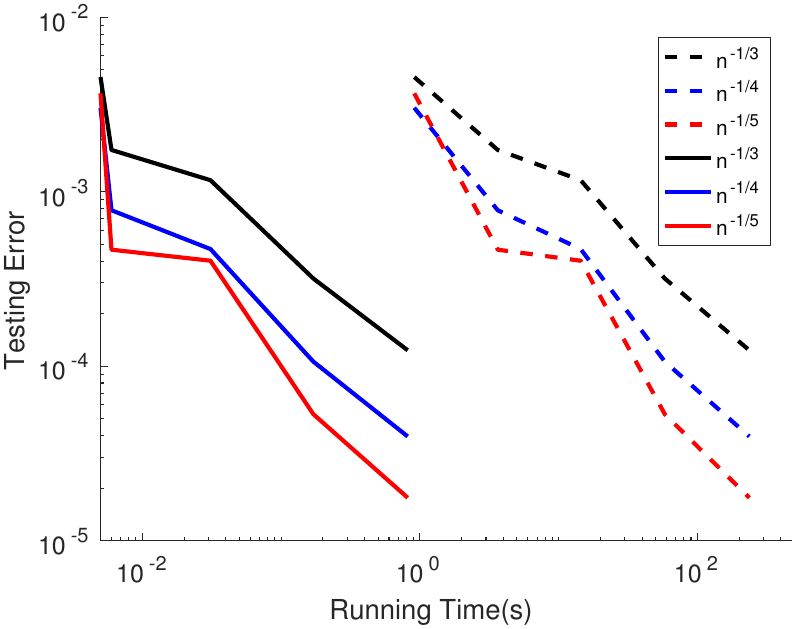}
		\caption{local quadratic regression ($k=2$)}
	\end{subfigure}
	\caption{Local polynomial regression for one dimensional data with running time constraints.  
		Solid lines correspond to FastLPR and dot lines correspond to BaseLine.
	}
	\label{fig:time_budget}
\end{figure*}

From Figures~\ref{fig:d1k0}, \ref{fig:d1k1} and \ref{fig:d2}, we observe that
as the number of testing points ($s$) or the training points become larger, the advantages of our proposed algorithm are more significant.
In particular, when 1,024,000 training and testing points are present under the setting of $k=1$ and $d=1$, Algorithm~\ref{alg:main} achieves $40,000\times$ speed-up
over the naive implementation in \texttt{locpol}.
The reason is that the naive implementation has $\Theta\left(ns\right)$ time complexity whereas ours is $O\left((n+s)\log^d n\right)$.

{
We compare our algorithm with a baseline implementation of local polynomial regression in Figures \ref{fig:d3k0}, \ref{fig:d3k1}, \ref{fig:d3k2}
on 3-dimensional data (i.e., $d=3$) with polynomial degrees ranging from $k=0$ to $k=2$.
Because the \texttt{locpol} package in R only supports $k=0$ for $d>1$, we compare our proposed algorithm with our own brute-force implementation
of local polynomial regression in C++.
As we can observe from the figures, our proposed algorithm still consistently outperforms the baseline implementation for all polynomial degrees
in the case of $d=3$.

We also compare memory consumptions between our method and the naive method in Table~\ref{tab:memory}.
For all experiments we fix $d=3$ and $s = 4000$ (we found different $s$ do not affect the memory consumption of both methods by much).
As we can observe from Table \ref{tab:memory}, our implementation consumes significantly more memory compared to baseline
because the multi-dimensional binary indexed trees is quite a memory-intensive data structure,
even with lazy memory allocation schemes in place.
Nevertheless, the highest amount of memory consumption (approximately 7GB for $d=3$, $k=2$ and $n=64,000$)
is still well within the limit of modern main memories on desktop or workspace computers.
Such memory consumption is worthy because of the significant improvement in running time our proposed algorithm brings.
}

%In Figure~\ref{fig:d2}, we compare Algorithm~\ref{alg:main} and the naive implementation on two dimensional data with $k=0$.\footnote{`locpol' package only support $k=0$ for two dimensional local polynomial regression.}
%Again, our method is significantly faster.

%Lastly, in Figure~\ref{fig:time_budget}, we test for a fixed running time, how to use as many as training samples as possible to achieve lower testing error.
Lastly, we report in Figure \ref{fig:time_budget} the function reconstruction errors of both FastLPR and the baseline method in \texttt{locpol} under fixed running time budgets.
We fix number of testing points to be $4000$ and the bandwidth parameter $h$ takes values of $n^{-1/3}$, $n^{-1/4}$ and $n^{-1/5}$.
Figure~\ref{fig:time_budget} shows under given time budgets, our proposed FastLPR algorithm processes much more observations (training samples)
and therefore achieves significantly smaller reconstruction error.

\section{Discussion}

\paragraph{\textbf{Kernel functions}.}
In our accelerated local polynomial estimation algorithm, the kernel $K(\cdot)$ is assumed to be the box kernel $K(u)=\mathbb I[|u|\leq 1/2]$
and for multiple dimensions ($d\geq 2$) the ``composite kernel'' is defined in terms of $\ell_\infty$ norm, meaning that the kernel evaluation between $x,z\in\mathbb R^d$
is $K(\|x-z\|_\infty/h)$.
The box kernel ensures that the sufficient statistics can be expressed as a linear combination of unweighted partial sums,
and the $\ell_\infty$ norm in multi-variate kernel evaluations lead to rectangle regions whose edges are parallel to the standard basis.
Both properties are crucial for our algorithmic development that enables fast sufficient statistics computation via binary indexed trees.

In general, box kernels with $\ell_\infty$ distance measures are sufficient as they achieve the same minimax statistical efficiency
as other kernel functions and/or equivalent distance measures.
However, for certain applications it might be desirable to consider kernel functions \emph{smoother} than the box kernel (e.g., the Gaussian kernel $K(u)=e^{-u^2/2}/\sqrt{2\pi}$)
to obtain smoother function fits.
We believe significantly different techniques are required to handle such kernels that are non-uniform and not restricted to parallel rectangular neighborhoods of testing points.

%\vspace{-0.1in}
\paragraph{\textbf{Incremental training and testing sets}.}
While in the description of our algorithm the sizes of both training and testing sets ($n$ and $s$) are fixed a priori,
we remark that our algorithm can perfectly handle training and testing data with increasing sizes.
In particular, for every new training or testing point, the corresponding update or interrogation paths are calculated in $O(\log^d n)$ time
and the Hash tables can be updated/queried using a similar amount of running time.
This property is particularly useful in applications where training data constantly grow/change, 
and estimates on incoming test points have to be made in a real-time fashion.

%\vspace{-0.1in}
\paragraph{\textbf{Further acceleration with approximate computation}.} 
In cases where exact computations of local polynomial estimates are not mandatory and small error in the estimates can be tolerated,
it is possible to further reduce the time and space complexity beyond $O(n\log^d n)$.

One approach is to consider significantly smaller (shorter) Hash tables with capacity $b\ll n\log^d n$.
Because the capacity of Hash tables is significantly smaller than the number of entries created by binary indexed trees, collisions are unavoidable.
Instead of resolving collisions completely, one can borrow the ideas of \textsc{CountSketch} \citep{charikar2004finding} that multiplies an additional Rademacher 
variable
\footnote{A Rademacher random variable takes on values of $\pm 1$ with equal probability.}
$\sigma(i_1,\cdots,i_d)$ in Hash table updates:
\begin{align*}
	\mathcal H_\ell(H(i_1,\cdots,i_d)) &\gets \mathcal H_\ell(H(i_1,\cdots,i_d)) \\
	&\;\;\;\; + \sigma(i_1,\cdots,i_d)T_\ell(i_1,\cdots,i_d).
\end{align*}

{
\paragraph{\textbf{Dependency on data dimension $d$}.}
The time complexity of our proposed algorithm has a leading $O((2k)^d)$ term which scales exponentially with data dimension $d$,
which is dropped in the main $O((n+s)\log^d n)$ time complexity bound because both polynomial degree $k$ and data dimension $d$ are treated
as constants in this paper.
On the other hand, the naive approach has an $O(n^2)$ time complexity and does not explicitly depend exponentially on $d$.
Nevertheless, in nonparametric regression/estimation it is typical that an exponential number of sample points are required 
(e.g., to estimate a Lipschitz continuous function up to precision $\varepsilon$, $\Omega(\varepsilon^{-(2+d)})$ samples are needed \citep{tsybakov2009introduction}).
Hence, our $O((n+s)\log^d n)$ time complexity is generally much more preferable than the $O(n^2)$ time complexity of naive approaches.
}

%\vspace{-0.2in}
\section{Conclusion}

In this paper we propose nearly linear time algorithms that compute local polynomial estimates for nonparametric density and regression function estimates.
Our algorithm is based on the novel application of multivariate binary indexed trees together with discretization and hashing techniques.
Simulation results demonstrate an up to $40000\times$ speed-up over state-of-the-art R implementation of local polynomial regression.
Future directions including general kernel functions and further algorithmic acceleration via sketching approximation are discussed.

\bibliographystyle{informs2014}
\bibliography{fastlpr}

% CASE 1: BiBTeX used to constantly update the references 
%   (while the paper is being written).
%\bibliographystyle{informs2014} % outcomment this and next line in Case 1
%\bibliography{<your bib file(s)>} % if more than one, comma separated

% CASE 2: BiBTeX used to generate mypaper.bbl (to be further fine tuned)
%\input{mypaper.bbl} % outcomment this line in Case 2

\end{document}